\documentclass[a4paper,fleqn,usenatbib]{mnras}

\usepackage{newtxtext,newtxmath}

\usepackage[T1]{fontenc}
\usepackage{ae,aecompl}

\usepackage{graphicx}	
\usepackage[]{subfig}
\usepackage{amsmath}	
\usepackage{soul}		
\usepackage{graphicx,rotating,booktabs}
\title[eFAM: application to SDSS-DR12 Combined Sample]
{Extended Fast Action Minimisation method: application to SDSS-DR12 Combined Sample}

\author[E. Sarpa et al.]{
	E. Sarpa,$^{1,2,3}$\thanks{E-mail: elena.sarpa@lam.fr}
	A. Veropalumbo,$^{2,3}$
	C. Schimd,$^1$
	E. Branchini,$^{2,3,4}$
	S. Matarrese.$^{5,6,7,8}$
	\\ 
$^{1}$ Aix  Marseille  Univ,  CNRS, CNES, LAM,  Marseille,  France \\
$^{2}$ Dipartimento di Matematica e Fisica, Universit\`a degli studi Roma Tre, Via della Vasca Navale, 84, 00146 Roma, Italy \\
$^{3}$ INFN - Sezione di Roma Tre, via della Vasca Navale 84, I-00146 Roma, Italy\\
$^{4}$ INAF - Osservatorio Astronomico di Roma, via Frascati 33, I-00040 Monte Porzio Catone (RM), Italy\\
$^{5}$ Dipartimento di Fisica e Astronomia ``Galileo Galilei'', Universit\`a degli studi di Padova, Via F. Marzolo, 8, I-35131 Padova, Italy \\
$^{6}$ INFN, Sezione di Padova, via F. Marzolo 8, I-35131, Padova, Italy\\
$^{7}$ INAF-Osservatorio Astronomico di Padova, Vicolo dell'Osservatorio 5, I-35122 Padova, Italy\\
$^{8}$ Gran Sasso Science Institute, Viale F. Crispi 7, I-67100 L'Aquila, Italy
}

\date{Accepted XXX. Received YYY; in original form ZZZ}

\pubyear{2018}

\begin{document}
\label{firstpage}
\pagerange{\pageref{firstpage}--\pageref{lastpage}}
\maketitle

\begin{abstract}
We present the first application of the extended Fast Action Minimization method (eFAM) to a real dataset, the SDSS-DR12 Combined Sample,
to reconstruct galaxies orbits back-in-time, their two-point correlation function (2PCF) in real-space, and enhance the baryon acoustic oscillation (BAO) peak. 
For this purpose, we introduce a new implementation of eFAM that accounts for selection effects, survey footprint, and galaxy bias. 

We use the reconstructed BAO peak to measure the angular diameter distance, $D_\mathrm{A}(z)r^\mathrm{fid}_\mathrm{s}/r_\mathrm{s}$, and the Hubble parameter, $H(z)r_\mathrm{s}/r^\mathrm{fid}_\mathrm{s}$, normalized to the sound horizon scale for a fiducial cosmology $r^\mathrm{fid}_\mathrm{s}$, at the mean redshift of the sample $z=0.38$, obtaining $D_\mathrm{A}(z=0.38)r^\mathrm{fid}_\mathrm{s}/r_\mathrm{s}=1090\pm29$(Mpc)$^{-1}$, and $H(z=0.38)r_\mathrm{s}/r^\mathrm{fid}_\mathrm{s}=83\pm3$(km~s$^{-1}$~Mpc$^{-1}$), in agreement with previous measurements on the same dataset.

The validation tests, performed using 400 publicly available SDSS-DR12 mock catalogues, reveal that eFAM performs well in reconstructing the 2PCF down to separations of $\sim 25h^{-1}\mathrm{Mpc}$, i.e. well into the non-linear regime.  
Besides, eFAM successfully removes the anisotropies due to redshift-space distortion (RSD) at all redshifts including that of the survey, allowing us to decrease the number of free parameters in the model and fit the full-shape of the back-in-time reconstructed 2PCF well beyond the BAO peak. Recovering the real-space 2PCF, eFAM improves the precision on the estimates of the fitting parameters. When compared with the no-reconstruction case, eFAM reduces the uncertainty of the Alcock-Paczynski distortion parameters $\alpha_{\perp}$ and $\alpha_{\parallel}$ of about $40$ per cent and that on the non-linear damping scale $\Sigma_\mathrm{\parallel}$ of about $70$ per cent.

These results show that eFAM can be successfully applied to existing redshift galaxy catalogues and should be considered as a reconstruction tool for next-generation surveys alternative to popular methods based on the Zel'dovich approximation.
\end{abstract}





\section{Introduction}
Baryon acoustic oscillations (BAO) are one of the main cosmological probes to investigate the nature of dark energy and search for deviations from General Relativity on cosmological scales. For this reason, they have been selected as primary science case of both current spectroscopic surveys, such as WiggleZ \citep{WiggleZ2009AAONw.115....3D}, the Baryon Oscillation Spectroscopic Survey \citep[BOSS;][]{Boss2013AJ....145...10D}, and its successor the extended Baryon Oscillation Spectroscopic Survey \citep[eBOSS;][]{eBOSS2017AJ....154...28B}, and future surveys operated by the Dark Energy Spectroscopic Instrument  \citep[DESI;][]{DESI_white_Aghamousa2016TheDE}, the ESA Euclid mission \citep{Euclid2011arXiv1110.3193L}, and the Roman Observatory \citep{WFIRST2012arXiv1208.4012G}. Thanks to their wide sky coverage, all the new-generation surveys allows the measurement of the acoustic scale with unprecedented accuracy. However, to reach the ambitious goal of sub-per cent precision and extract the maximum information from the non-linear scales, one needs to pair the quality of observations with an accurate model of the small-scale clustering. To this end, the use of highly-optimized BAO reconstruction techniques able to minimise the non-linear effects that blur the acoustic feature and possibly trace the orbits of galaxies backwards-in-time is nowadays an essential step of any clustering analyses.

For the official SDSS-III/BOSS analysis, the Zel'dovich reconstruction technique \citep[ZA;][]{Eisenstein2007a,Padmanabhan+2012} has been successfully adopted, yielding a remarkable improvement on the BAO measurements in both momentum \citep{Beutler2017MNRAS.464.3409B} and configuration space \citep{Alam:2016hwk}. However, the intrinsic linearity of the method prevents the accurate modelling of redshift-space distortions (RSD), usually overestimating their amplitude \citep{kazin2013clustering}, and limits the range of validity of the recovered linear two-point correlation to relatively large separations \citep{PadmanabhanWhiteCohn2009}. Pursuing a more accurate description of the real-space linear correlation function, new non-linear BAO reconstruction techniques have been proposed. In particular, iterative reconstruction techniques \citep[e.g.][]{Schmittfull2017,Hads2018,Wang2017}) have attracted special attention and have been successfully tested on $N$-body dark matter simulations. \cite{Mao2020} proposed an innovative approach which intends to recover the BAO signal using deep convolutional neural networks. The extended Fast Action Minimization method \citep[eFAM;][]{Sarpa:2018ucb}, used in this work, is a variational reconstruction method designed to recover the geodesics of the mass tracers via minimization of the action of the system.

In this study, we present the first application of eFAM to spectroscopic data emphasising its distinctive capability to simultaneously recover the real-space correlation function at both the observed and high redshift well into the mildly non-linear regime. To fully exploit the non-linearity of eFAM, we apply the reconstruction on the lower redshift bin $0.2<z<0.55$ extracted from the BOSS-DR12 Combined Sample, which is mostly affected by non-linear clustering. Finally, we measure the expansion rate $H(z)$ and the angular diameter distance $D_\mathrm{A}(z)$ before and after reconstruction using the clustering wedges \citep{Kazin:2011xt}.

The paper is organized as follows. In Section~\ref{sec:BAOreconstruction} we illustrate the new features of the eFAM algorithm accounting for masked regions, fibres collision, flux selection, and galaxy bias, as required for the application to wide-field, massive spectroscopic galaxy surveys. The BOSS-DR12 data and the mocks employed to test the method and build the covariance matrix are illustrated in Section~\ref{sec:data+mocks}. In Section~\ref{sec:Metodology} we describe the statistical tools used in the analysis, and motivate the choice of the fiducial fitting procedure in Section~\ref{2PCFmodel}. We study the systematic behaviour of eFAM by reconstructing 400 realistic mocks in Section~\ref{sec:Results}, and analyse SDSS data in Section~\ref{sec:ResultsData}. In Section \ref{sec:Discussion} we discuss the main points of our study and present our conclusions.

\begin{figure*}
\centering
\includegraphics[width=0.95\textwidth]{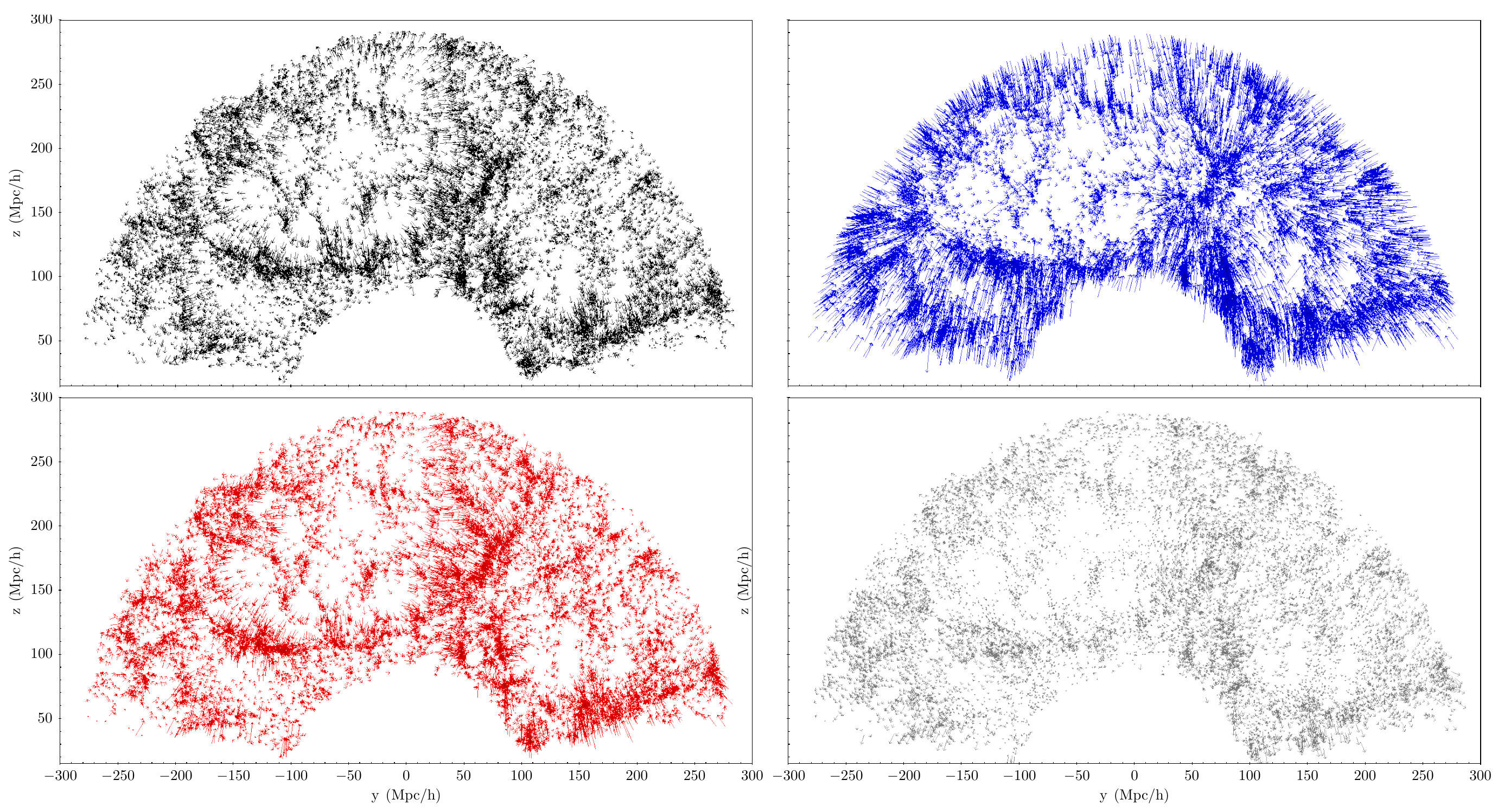} 
\caption{2D velocity maps for the BOSS-like sub-sample carved from the \textsc{deus} dark matter halos simulation. {\it Top-left panel:} N-body velocity field. {\it Top-right panel:} velocity field reconstructed without accounting for the external tidal field and assuming an empty universe outside the survey; the bulk flow induced by the geometry of the survey is dominant. {\it Bottom-left panel:} velocity field reconstructed including the external tidal field; the geometrical bulk flow is successfully removed and the N$_\mathrm{body}$ velocity field efficiently recovered. {\it Bottom-right panel:} residuals velocity field V$_\mathrm{Nbody}$-V$_\mathrm{eFAM}$ after correcting for the external tidal field.}
\label{fig:tidal_corr}
\end{figure*}

\section{\lowercase{e}FAM for real galaxy surveys} \label{sec:BAOreconstruction}

The eFAM method, described in details in \cite{Sarpa:2018ucb}, reconstructs the full-orbit of $N$ point-like mass particles at their observed positions
$\{\mathbf{x}_i(t)\}_{i=1,\cdots, N}$, by minimizing the action of the self-gravitating system
\begin{eqnarray}\label{eq:action}
    S&=&\sum_{i=1}^N\int_0^{D_\mathrm{obs}}\mathrm{d}D\left[fEDa^2\frac{1}{2}\left(\frac{\mathrm{d}\mathbf{x}_i}{\mathrm{d}D}\right)^2 \right. \nonumber \left.+\frac{3\Omega_{\mathrm{m},0}}{8\pi fEDa}\phi_\mathrm{tot}(\mathbf{x}_i)\right] \nonumber \\
     && +\sum_{i=1}^N\frac{1}{2}(fEDa)_\mathrm{obs}^2\left(\frac{d\mathbf{x}_{i,\mathrm{obs}}}{dD}\cdot\mathbf{x}_{i,\mathrm{obs}}\right)^2
\end{eqnarray}
using mixed boundary conditions \citep{peebles1989tracing}. Under the hypothesis that galaxies trace the underling mass distribution, boundary conditions are set by the galaxies observed redshifts, $\mathbf{s}_{i,\mathrm{obs}}$, and their assumed vanishing peculiar velocities at early times.
Here, $a$ indicates the scale factor, $D$ is the linear growth factor that we used as time variable, 
$f={ \rm d} \ln{D} / {\rm d} \ln{a}$ is the linear growth rate,
and $E = H/H_0$ is the dimensionless Hubble parameter. $\Omega_{\rm m}$ represents the mass density parameter and the subscripts `0' and `obs' indicate, respectively, quantities measured at the present epoch and at the redshift of the survey $z = z_\mathrm{obs}$. Following \cite{nusser2000least}, we model galaxies' orbits as a linear combination of $M$ time-dependent basis functions $\left\{q_n(D)\right\}_n$ viz.
\begin{equation}
\mathbf{x}_i(D)=\mathbf{x}_{i,\mathrm{obs}}+\sum_{n=0}^M\mathbf{C}_{i,n}q_n(D)
\end{equation}
and minimise the action in Equation~(\ref{eq:action}) with respect to the expansion coefficients $\mathbf{C}_{i,n}$.

In this work, we modify the original implementation of the eFAM algorithm \citep{Sarpa:2018ucb}, which has been already tested on simulated halos catalogues, to optimize its application to real surveys. Specifically, we focus on the inclusion of unique selection effects, observational biases, and sample geometry in the computation of the total gravitational potential of the system, $\phi_\mathrm{tot}$.


\vspace{-0.4cm}
\subsection{Survey geometry}\label{subsec:tidal}
Real surveys probe the mass distribution within a large yet finite region of the Universe. The total gravitational potential at any point $\mathbf{x}$ in this region, $\phi_\mathrm{tot}(\mathbf{x})$, is given by the sum of two terms: the potential due to the matter distribution inside the volume of the survey, $\phi_\mathrm{int}(\mathbf{x})$, and the potential due to the external mass distribution, $\phi_\mathrm{ext}(\mathbf{x})$, which is unknown. Neglecting the external matter distribution in the estimation of $\phi_\mathrm{tot}$ in Equation~(\ref{eq:action}) generates spurious tidal fields that affect the quality of the reconstruction. To mitigate their effect, a typical approach is to estimate $\phi_\mathrm{ext}(\mathbf{x})$ assuming a homogeneous and isotropic distribution of matter outside the survey volume with the same mean density of the galaxy sample. This procedure removes the spurious tidal fields due to the geometry of the survey. However, since we do not model the clustering properties of the external mass distribution, residual tidal effects are expected to degrade the quality of the reconstruction as we approach the edge of the survey volume. For this reason, we define a buffer region at a distance $d_\mathrm{Buffer}$ from the edges that we use to perform the reconstruction but discard in the analysis of the reconstructed field.
The optimal value for $d_\mathrm{Buffer}$ depends on the survey geometry and on the desired accuracy of the reconstruction and thus needs to be evaluated in each specific eFAM application. 

A simple, brute force implementation of this strategy would be to embed the survey region in a much larger volume filled with an un-clustered distribution of synthetic objects with same angular and radial selection function and mean density as the real galaxies of the survey, and run eFAM over all objects, real and synthetic alike. However, the computational time required by this procedure is prohibitive since the number of synthetic objects needs to be very large to minimize shot-noise errors (with the Poisson solver implemented by eFAM, the CPU time required to compute the gravitational potential increases linearly with the number of objects).

For this reason, we adopt instead an equivalent, more effective approach based on the Newton's shell theorem: \emph{i)} we assume a homogeneous isotropic distribution of mass throughout the whole Universe, inside and outside the survey volume, and set $\phi_\mathrm{tot}(\mathbf{x})=0$ everywhere; \emph{ii)} we use a Poisson solver to numerically estimate the potential $\phi_\mathrm{int}$ as generated by a uniform distribution of matter within the survey volume; \emph{iii)} we set $\phi_\mathrm{ext}=-\phi_\mathrm{int}$. As mentioned above, the CPU time required for the computation of the potential due to synthetic objects is larger than the one needed to compute it from galaxies. Though, here the calculation is performed only once per eFAM reconstruction and not at each time-step in Equation~(\ref{eq:action}). The increase in computational time is, therefore, negligible.

To assess the accuracy of this procedure, we apply eFAM to a set of dark matter halos extracted from the \textsc{deus} simulation \citep{Rasera+2014} with same footprint and similar depth as the BOSS-DR12 survey. Figure \ref{fig:tidal_corr} illustrates the comparison between the ``true'' velocity field in the $N$-body simulation (top-left panel) and the reconstructed one before (top-right panel) and after (bottom-left panel) including the effect of the model tidal field $\phi_\mathrm{ext}$. Our procedure effectively removes the spurious bulk-flow that would otherwise dominate the velocity field. The velocity residuals map (bottom-right panel) confirms that spurious flows are confined to the regions near the edges justifying our choice of discarding those regions in the analysis. A quantitative assessment of the quality of the reconstruction and a test designed to evaluate systematic errors is illustrated in Appendix~\ref{appendix:Survey_geom}.

\vspace{-0.4cm}
\subsection{Galaxy bias}\label{subsec:bias}
The original implementation of FAM \citep{nusser2000least} assumes that the average mass density of the survey volume matches the cosmological mean and that all the mass is associated to individual, discrete and visible objects. However, real galaxies are known to be biased tracers of the mass density field. To account for galaxy bias, we assume that bias is a local, linear and deterministic phenomenon and that the total mass distribution can be split in two components: a distribution of point-like discrete masses accounting for both the luminous and dark matter associated to each observed galaxy, $\rho'_\mathrm{g}(\mathbf{x})$, surrounded by a uniform, smooth distribution of dark matter, $\bar{\rho}_\mathrm{DM}$, filling the survey volume, i.e.
\begin{equation}
\rho_{\mathrm{tot}}(\mathbf{x})\equiv \rho'_\mathrm{g}(\mathbf{x})+ \bar{\rho}_\mathrm{DM}\sum_\mathrm{g}\left[1-\delta_\mathrm{D}(\mathbf{x}-\mathbf{x}_\mathrm{g})\right],
\end{equation}
where the sum runs over galaxy positions $\mathbf{x}_\mathrm{g}$.
The relation between the observed galaxy density field, $\rho_\mathrm{g}$, and $\rho'_\mathrm{g}$ is set by the linear biasing assumption $\delta_\mathrm{tot}=\delta_\mathrm{g}/b$ to
\begin{equation}\label{eq:rho_tot}
 \rho'_\mathrm{g}(\mathbf{x})=\left(\frac{\delta_\mathrm{g}(\mathbf{x})}{b}+1\right)\Omega_\mathrm{m,0}\frac{3H^2}{8\pi G}\sum_\mathrm{g}\delta_\mathrm{D}(\mathbf{x}-\mathbf{x}_\mathrm{g}),
\end{equation}
where $\Omega_\mathrm{m,0}3H^2/\left(8\pi G\right)=\bar{\rho}_\mathrm{tot}$, $\delta_\mathrm{tot}=1+\rho_\mathrm{tot}/\bar{\rho}_\mathrm{tot}$, and $\delta_\mathrm{g}=1+\rho_\mathrm{g}/\bar{\rho}_\mathrm{g}.$

Equation~({\ref{eq:rho_tot}}) implies that the net effect of the bias is to down-weight the mass of individual objects, hence lowering their contribution to the mass density below the cosmological mean. Since eFAM solely accounts for the mass component associated with discrete objects, the down-weighting of the masses will cause the survey volume to be treated as an underdense region and a net outflow will be predicted. To restore the density balance, we thus need to account for the gravitational pull of the smooth dark component. Our strategy is to model the gravitational potential generated by the smooth dark matter distribution, $\phi_\mathrm{smooth}$, as an extra unknown external field to be added to the one exerted by the homogeneous external distribution described in Section~\ref{subsec:tidal}.  
Similarly to the procedure adopted for the external tidal field, we infer the new external gravitational potential $\phi'_\mathrm{ext}\equiv \phi_\mathrm{ext}+\phi_\mathrm{smooth}$ setting $\phi'_\mathrm{ext}=-\phi'_\mathrm{int}$, where $\phi'_\mathrm{int}$ is now estimated having fixed the mean density of the smooth matter distribution within the survey to  $\bar{\rho'}_\mathrm{R}=\bar{\rho}_\mathrm{tot}-\bar{\rho}_\mathrm{DM}$.

\vspace{-0.2cm}
\section{Data and simulations}\label{sec:data+mocks}

\subsection{SDSS-DR12 Combined Sample}\label{subsec:data}
We apply eFAM reconstruction algorithm to the galaxies in the low-redshift bin $0.2<z<0.55$ of the ``BOSS Combined Sample'', corresponding to the first redshift bin considered in the clustering analyses of the BOSS collaboration  \citep{kitaura2016clustering,Ross:2016gvb,Vargas-Magana:2016imr}. The ``Combined Sample'' spans the redshift range $0.2<z <0.75$ and includes both the Constant Stellar Mass sampe (CMASS), designed to be approximately stellar-mass limited above $z = 0.45$, and the low-redshift (LOWZ) sample, targeting galaxies in the redshift range $0.15 < z <0.43$. Having considered the survey footprint, we decide to focus on the northern galactic cap to reduce both the computational cost of the reconstruction and the impact of the external tidal field, which would be large in the thinner southern strip. Following \cite[][hereafter Ross17]{Ross:2016gvb}, we assign a statistical weight to each galaxy to account for fibre collision, veto flag, and the need to minimize the variance of the clustering estimator. The latter aspect is dealt with using the so-called FKP weight \citep{Feldman:1993ky} $\omega_\mathrm{FKP}=1/[1+\bar{n}(z_i)P(k_0)]$ where $\bar{n}$ is the redshift distribution at the galaxy redshift $(z_i)$, and $P(k_0)$ = $40000(h^{-1}$Mpc$)^3$ is the amplitude of the power spectrum evaluated approximately at the BAO scale for SDSS Luminous Red Galaxies (LRG). In the eFAM implementation, we use these weights to assign an effective mass to each individual object.

\vspace{-0.4cm}
\subsection {MultiDark-Patchy mock catalogues}\label{subsec:mocks}

To test the performances of eFAM and estimate the covariant errors employed in the likelihood analysis, we make use of the publicly available MultiDark-Patchy (hereafter MD-Patchy) mocks \citep{kitaura2016clustering, rodriguez2016clustering}. MD-Patchy are extracted from the Big-MultiDark Planck simulation \citep{klypin2016multidark} which assumes a spatially flat $\Lambda$CDM $Planck$ cosmology with $\Omega_\mathrm{m}$ = 0.307115, $\Omega_\Lambda$ = 0.692885, $\Omega_\mathrm{b}$ = 0.048, $\sigma_8$ = 0.8288, $n_\mathrm{s}$= 0.9611, $h$ = 0.6777. Furthermore, a non-linear, deterministic, stochastic bias recipe has been applied to the galaxy catalogue to match the clustering properties of the BOSS LRG sample.

In the following analysis, we use the V6-C version of the ``Combined'' sample mocks, built to estimate the covariance matrix of the 1, 2, and 3-points clustering statistic of SDSS-DR12/BOSS. Since we are interested in computing the 2PCF of the mock galaxies, we also use the associated mock random catalogue which contains 50 times as many objects as the SDSS-DR12 mocks and has the same footprint, selection function, and statistical weights. To reduce the computational cost of our analysis, we decide to consider only 400 mocks out to the 1000 employed by Ross17. The impact of this choice is discussed in Section~\ref{sec:Discussion}.
\vspace{-0.2cm}
\section{Statistical tools}\label{sec:Metodology}
\begin{figure*}
\centering
\includegraphics[width=0.95\textwidth]{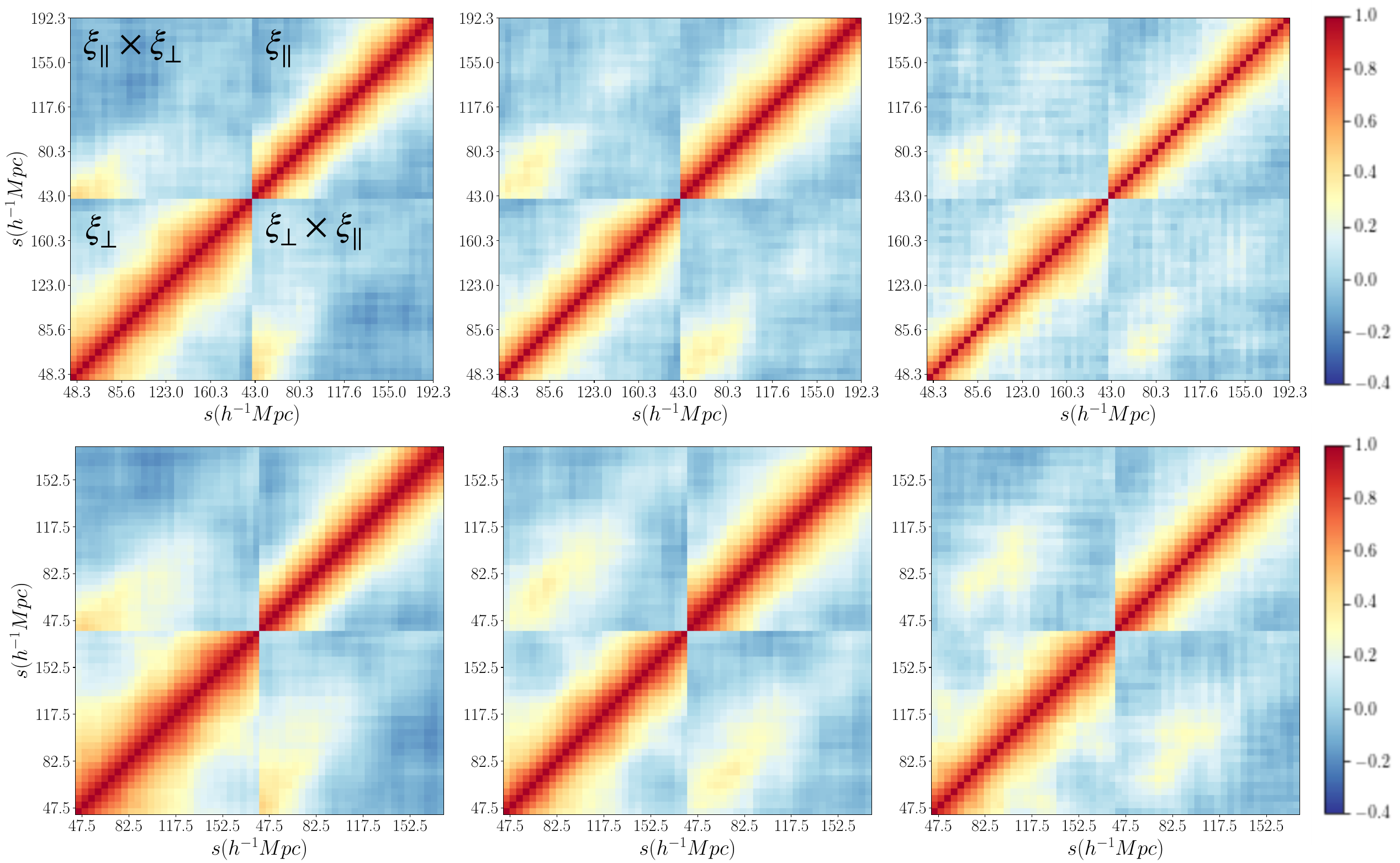}
\caption{Correlation matrices of the clustering wedges estimated from 400 MD-Patchy mocks. \emph{Left panels}: Obs sample. \emph{Middle panels}: RecZ samples. \emph{Right panels} RecL samples. 
\emph{Top panels:} catalogs extracted by excluding all objects laying within $d_\mathrm{Buffer}=200h^{-1}$Mpc from the survey edges. \emph{Bottom panels:} $d_\mathrm{Buffer}=125h^{-1}$Mpc. The color code is set accordingly to the normalized covariance amplitude. 
}\label{fig:Correlation_matrix}
\end{figure*}

\subsection{Two-point correlation function estimator}\label{subsec:estimator}
To gauge the performance of the eFAM method, we perform a similar clustering analysis to that already carried out by the BOSS collaboration aimed at estimating the Hubble parameter, $H$, and the diameter angular distance, $D_\mathrm{A}$, from the analysis of the BAO peak in the anisotropic 2PCF, $\xi(s,\mu)$. Here, $s$ represents the separation of the galaxy pair and $\mu$ is the cosine of the angle between $\mathbf{s}$ and the bisector of the angle to the pair. To estimate the $\xi(s,\mu)$ of each sample,we use the minimum variance, unbiased \citet{LandySzalay1993} estimator
\begin{equation}\label{eq:LS}
    \xi(s,\mu)=\frac{DD(s,\mu)-2DR(s,\mu) +RR(s,\mu)}{RR(s,\mu)},
\end{equation} 
where the data-data, $DD$, data-random, $DR$, and random-random, $RR$, counts are evaluated in bins of $\Delta s=5h^{-1}$Mpc and $\Delta\mu=0.05$, within the ranges $30<s<190$ and $0<\mu<1$, respectively. We measure the 2PCF of three types of datasets: \emph{i}) the observed, redshift-space (hereafter labelled `Obs'), in which SDSS galaxies are placed at their comoving coordinates using measured redshift as distance proxy and assuming the cosmology of the mocks as fiducial model, \emph{ii}) the reconstructed real-space galaxies at the epoch of observation (labelled `RecZ'), obtained
by placing galaxies at their reconstructed real-space positions,
and \emph{iii}) the back-in-time reconstructed real-space positions (`RecL') in which galaxies are displaced back-in-time to the real-space positions they occupied at an earlier epoch.  
Unlike the standard ZA reconstruction, eFAM does not require any spatial smoothing of the observed density field to obtain the reconstructed one. We can thus use the estimator in Equation~(\ref{eq:LS}) rather than the modified version of \cite{Padmanabhan+2012}, which additionally requires to reconstruct the back-in-time orbits of the random objects, with a consequent important computational load.

To minimize the impact of the external tidal field, we estimate the 2PCF of the sample after having discarded all the objects, real and random alike, that reside in the buffer region (see Section~\ref{subsec:tidal}).
Besides, since reconstruction displaces the objects from their original positions, we trim the reconstructed catalogs to match the survey footprint and depth.
Finally, the 2PCF is estimated by assigning to each object its statistical weight accounting for FKP, fibre collision, and veto flag. 

\vspace{-.2cm}
\subsection{Clustering wedges}\label{subsec:wedges}
We analyse the anisotropic 2PCF focusing on the wedge clustering statistics introduced by \cite{Kazin:2011xt} and applied by \cite{kazin2013clustering} to the study of the SDSS-DR11 galaxy clustering. We shall consider the two wedges that model clustering along the parallel ($\parallel$) and transverse ($\perp$) direction to the line-of-sight:
\begin{equation}
\xi_{\parallel(\perp)}(s)=\frac{1}{\Delta \mu}\int_{0.5(0)}^{1(0.5)}d\mu \xi(s,\mu).
\label{eq:wedges}
\end{equation} 
Following \cite{kazin2013clustering}, we focus on the BAO peak whose scale depends on a combination of $H$, $D_\mathrm{A}$ and the sound horizon scale during baryon drag epoch, $r_\mathrm{s}$. Since the radial clustering wedge $\xi_\parallel$ is mostly sensitive to $H$ while the transverse one, $\xi_\perp$, is sensitive to $D_\mathrm{A}$, one can measure $\xi_\parallel$ and $\xi_\perp$ to break this degeneracy.
This is done by exploiting the Alcock-Paczynski effect \citep{Alcock1979Natur.281..358A} (hereafter AP), which quantifies the dependence of the measured separations along and across the line of sight, $s^\mathrm{fid}_\parallel$ and $s^\mathrm{fid}_\perp$, on a fiducial cosmological model, i.e.
\begin{eqnarray}
s^\mathrm{t}_\perp&=&\alpha_\perp s^\mathrm{fid}_\perp \quad\quad \mathrm{with} \quad\quad \alpha_\perp \equiv \frac{D^\mathrm{t}_\mathrm{A}}{D^\mathrm{fid}_\mathrm{A}}\frac{r^\mathrm{fid}_\mathrm{s}}{r^\mathrm{t}_\mathrm{s}}\label{eq:alpha_perp},\\
s^\mathrm{t}_\parallel&=&\alpha_\parallel s^\mathrm{fid}_\parallel\quad\quad \mathrm{with} \quad\quad \alpha_\parallel\equiv \frac{H^\mathrm{fid}}{H^\mathrm{t}}\frac{r^\mathrm{fid}_\mathrm{s}}{r^\mathrm{t}_\mathrm{s}},\label{eq:alpha_par} 
\end{eqnarray}
to infer the `true' values $H^{\mathrm{t}}$ and $D^{\mathrm{t}}$ from the modelling of the anisotropy in the measured 2PCF. The discrepancy between the fiducial and true separation, and thus between the fiducial and true cosmology, is quantified by the parallel and perpendicular dilation parameters, $\alpha_\parallel$ and $\alpha_\perp$.

In redshift-space, anisotropies in the 2PCF are generated by both RSD and AP-distortions. Hence, one needs to disentangle the two effects to accurately estimate $H^{\mathrm{t}}$ and $D^{\mathrm{t}}$.
In this work, we use eFAM to remove RSD. However, since the reconstruction algorithm demands to set a value for the linear growth rate, $f$, we shall not attempt to estimate its value in the likelihood analysis of reconstructed samples. Instead, we will check the adequacy of the fiducial $f$ by verifying that the RSD have been consistently removed in both reconstructed datasets, RecZ and RecL.

In addition to clustering wedges, we study the anisotropy in the 2PCF considering its multipoles 
  \citep{Padmanabhan:2008ag} 
\begin{equation}
    \xi_l(s)=\frac{2l+1}{2}\int_{-1}^1 d\mu\xi(s,\mu)\mathcal{L}_l(\mu),
\end{equation}
where $\mathcal{L}_l(\mu)$ are the Legendre polynomials, which provide an equivalent analysis of the clustering properties to that of the wedges but ease the estimation of anisotropies as a non-vanishing quadrupole.

\vspace{-.5cm}
\subsection{Covariance matrix}\label{subsec:cov}
Given the strong correlation between $\xi_\perp$ and $\xi_\parallel$, the covariance matrix of the clustering wedges $C_\mathrm{s}$ is built by combining the auto-covariance and the cross-covariance of the binned parallel and perpendicular wedges measured from the 400 mocks. As shown by \cite{hartlap2007your,percival2014clustering}, the precision matrix, $\Psi$, estimated by inverting the measured covariance matrix, $C_\mathrm{s}$, is a biased estimate of the true one, with the amplitude of this bias depending on both the number of mocks $n_\mathrm{s}$ used to measure $C_\mathrm{s}$ and its dimension 
$n_\mathrm{b} \times n_\mathrm{b}$. To correct for this bias we set
\begin{equation}\label{eq:hartlap_cov}
   \Psi=\left( 1-\frac{n_\mathrm{b}+1}{n_\mathrm{s}-1}\right)C_\mathrm{s}^{-1}.
\end{equation}
The core of our analysis focuses on the properties of the 2PCF in the separation range $[50,150]h^{-1}\mathrm{Mpc}$. For this application, we sample the clustering wedges in 20 bins of $5h^{-1}$Mpc covering the whole range of interest.

Figure \ref{fig:Correlation_matrix} illustrates the auto and cross-correlation matrices of the clustering wedges measured for the Obs (left panels), RecZ (middle) and RecL (right) mock samples. The difference between the top and bottom panels is due to the different extent of the buffer region that contains objects excluded from the analysis.
A simple visual inspection reveals that the application of eFAM significantly reduces the amplitude of the off-diagonal terms in both RecZ and RecL samples.  

\vspace{-0.4cm}

\section{Modelling the 2PCF}\label{2PCFmodel}
\begin{table*}
	\centering
	\caption{Fitting parameters in the likelihood analysis and their prior range. \emph{Top:} 
	parameters used to fit the correlation function averaged over the mocks $\langle\xi\rangle$ (Section~\ref{subsec:real_xi}). \emph{Bottom:} parameters used to fit the 2PCF of the data (Section \ref{sec:ResultsData} and that of a single mock (Section~\ref{subsec:AcoustiMocks}).}
	\label{tab:par_space}
	\begin{tabular}{lcccccccc}
		\hline
		sample & $\alpha_\perp$ &  $\alpha_\parallel$ &$\Sigma_\perp$ $(h^{-1}\mathrm{Mpc})$ &  $\Sigma_\parallel$ $(h^{-1}\mathrm{Mpc})$ &$b$ &$f$& $\Sigma_\mathrm{s}$ $(h^{-1}\mathrm{Mpc})$ & $\left\{A_{p,i}\right\}_{i=0,1,2}$ \\ 
		\midrule
		\midrule
		\multicolumn{9}{c}{$\langle\xi\rangle$ (Section~\ref{subsec:real_xi})} \\
		
		\hline 
     	All & Flat: $[0.7,1.3]$ & Flat $[0.7,1.3]$ & Flat: $[0,15]$ & Flat: $[0,15]$ &  Flat: $[0,10]$ &  Flat: $[0,5]$ & Flat: $[0,10]$ &  Flat: $[-10,10]$  \\
		
		\midrule
		\midrule
		\multicolumn{9}{c}{Data and individual mocks (Section~\ref{subsec:AcoustiMocks} and
	\ref{sec:ResultsData})} \\
		\hline 
	
		Obs &  Flat: $[0.7,1.3]$ &  Flat: $[0.7,1.3]$ & Fixed to $\langle\xi\rangle$ & Fixed to $\langle\xi\rangle$ & Fixed to $\langle\xi\rangle$ & Fixed to $\langle\xi\rangle$ & Fixed to $\langle\xi\rangle$ & Flat: $[-10,10]$  \\
		
		RecL & Flat: $[0.7,1.3]$ &  Flat: $[0.7,1.3]$ & Fixed to $\langle\xi\rangle$ & Fixed to $\langle\xi\rangle$ & Fixed to $\langle\xi\rangle$ &Fixed to 0 & Fixed to 0  & Flat: $[-10,10]$  \\

	    \bottomrule
	 \end{tabular}
\end{table*}
\subsection{Two-point correlation function model} \label{sec:ximodel}
The goal of this study is to model the anisotropic 2PCF over a large range of scales including the BAO peak. Hence, we need to take into account RSD, AP-distortions and non-linear clustering evolution. The first step is to model the non-linear mass power spectrum in real-space, $P^\mathrm{R}(k,\mu)$, using the \cite{EisensteinHu1998} model
\begin{align}
   P^\mathrm{R}(k,\mu)=\left[P_\mathrm{l}(k)-P_\mathrm{nw}(k)\right]\mathrm{e}^{-k^2\left[(1-\mu^2)\Sigma^2_\perp/2+\mu^2\Sigma^2_\parallel/2 \right]}+P_\mathrm{nw}(k),\nonumber
\end{align} where $P_\mathrm{l}(k)$ is the linear power spectrum, $P_\mathrm{nw}$ is the no-wiggle power spectrum,
and $\Sigma_\parallel$ and $ \Sigma_\perp$ quantify the non-linear damping along and across the line-of-sight \citep{CrocceScoccimarro2006,Eisenstein:2007b, CrocceScoccimarro2008}.
$P_\mathrm{l}$ is estimated using CAMB \citep{CAMB2000ApJ...538..473L} having assumed the same cosmological model as the mock catalogs while $P_\mathrm{nw}$ is computed analytically.

To model RSD, we combine linear theory to model large-scale coherent motion \citep{Kaiser1987}, with the streaming model \citep{FOG1994MNRAS.267.1020P} to account for small-scale incoherent velocities obtaining
\begin{equation}\label{eq:pk_rsd}
  P^\mathrm{S}(k,\mu)=\frac{\left(1+\mu^2f/b\right)^2}{1+(k^2\mu^2\Sigma^2_s)^2}P^\mathrm{R}(k,\mu) \, ,
\end{equation}
where the superscript `S' indicates quantities measured in redshift-space, $\mu$ is the cosine angle between the wave vector and the line-of-sight direction, $b$ is the linear bias parameter, and $\Sigma_\mathrm{s}$ is the velocity dispersion parameter. The redshift-space template is only used to fit the 2PCF of the Obs sample, i.e. before performing eFAM reconstructions; it is not included to model the reconstructed 2PCF of the RecZ and RecL sample since we find that the reconstructed 2PCF of the RecZ and RecL samples is well fit by a model with $f=\Sigma_\mathrm{s}=0$ (see Section~\ref{subsec:real_xi}).

Finally, we model the measured clustering wedges, $\{\tilde{\xi}^\mathrm{R(S)}_p\}_{p=\perp,\parallel}$, as follows: \emph{i)} we estimate the multipoles of the anisotropic power spectrum, $P^\mathrm{R(S)}(k,\mu)$, defined as
\begin{equation}
P^\mathrm{R(S)}_l(s)=\frac{2l+1}{2}\int_{-1}^1d\mu P^\mathrm{R(S)}(k,\mu)
\mathcal{L}_l(\mu);
\end{equation}
\emph{ii)} we compute the corresponding 2PCF multipoles $\xi_l$ performing a Fourier transform of each multipole $P_l$; \emph{iii)} we combine the 2PCF multipoles to obtain the clustering wedges, $\xi_p^\mathrm{R(S)}(s)$ as described in \cite{Kazin:2011xt}, and \emph{iv)} we obtain the fiducial template by adding the full-shape term $A_p(s)$, viz.
\begin{equation}\label{eq:xi_temp}
    \tilde{\xi}^\mathrm{R(S)}_p(s;\alpha_\perp, \alpha_\parallel)=\xi_p^\mathrm{R(S)}(s;\alpha_\perp,\alpha_\parallel)+A_p(s).
\end{equation}
Here, the two dilation parameters are free parameters accounting for the mismatch between the fiducial and the true cosmology while the full-shape term $A_p(s)=A_{p,0}+A_{p,1}/s+A_{p,2}/s^2$ is introduced to account for possible systematic errors in reconstructing the broad-band shape of the 2PCF. Differently from the fiducial BOSS template (Ross17), we omit the shape parameter $B$ that would be highly degenerate with those controlling the RSD.

\vspace{-.2cm}
\subsection{Likelihood analysis}
\subsubsection{Maximization of the likelihood}
To fit the model to the measured 2PCF, we search for the maximum
of the likelihood in the space of the free parameters using the modelling routine in the CosmoBolognaLib library \citep{CosmoBolognaLib2016A&C....14...35M}. 
Maxima are searched for using a two-step procedure. First, the Nelder-Med method \citep{nelder1965simplex} is used to approach the best-fit value. Here, the the $N_\mathrm{p}$-dimensional parameters space is probed by evaluating the likelihood at the vertex of a running $(N_\mathrm{p}+1)$-dimensional simplex which progressively approaches the nearest maximum. Second, once convergence is attained, we refine the search by running a Markov Chain around the best-fit value. 
In the following, we perform different likelihood analyses varying the number of free parameters. Table~\ref{tab:par_space} summarizes our choices for the analysis of the mean correlation function averaged over 400 mocks, $\langle \xi \rangle$, and for the study of individual mocks.

Similarly to the majorities of the multidimensional optimization algorithms, the Nelder-Med method approximates a local rather than the global maximum. To ensure convergence to the global maximum, it is crucial to set the initial size of the simplex so as to cover a sufficiently large portion of the $N_\mathrm{p}$-dimensional domain. We do so by starting the maximization of the likelihood from an equilateral simplex and re-scaling the fitting parameters so that to have similar magnitudes. The parameters $\alpha_{\perp,\parallel}$, $\Sigma_{\perp,\parallel}$, $\Sigma_\mathrm{s}$, $f$, and $b$ represents physical quantities thus their amplitude can be guessed from perturbation theory \citep{CrocceScoccimarro2008}. Ross17 estimate them to be in the range $[1,10]$. On the contrary, the values of the shape parameters $\left\{A_{p,i} \right\}_i$ is {\it a priori} unknown and their values can significantly vary. To avoid handling quantities with very different magnitudes we normalize the values of $A_{p,i}$ by the factor $3\sigma_{\xi_p(r_\mathrm{ref})}r_\mathrm{ref}^i$, where $\sigma_{\xi_p(r_\mathrm{ref})}$ is the standard deviation of $\xi_p(r_\mathrm{ref})$ among the mocks evaluated at an arbitrary scale $r_\mathrm{ref}$. With this parameterization, and assuming $A_{p,i}=0 \;\forall i$ for the mean wedges averaged over all the mocks $\langle\xi_p \rangle$, we are able to model the 2PCF at $r_\mathrm{ref}$ for the almost totality of the mocks using $|A_{p,i}| \lesssim 1$. Aiming at modelling the distortion of the 2PCF that can not be described solely by physical parameters (e.g. a negative value of the correlation function at small scales), we set $r_\mathrm{ref}\sim 80h^{-1}$Mpc, corresponding to the deep of the correlation function at small separations. 

\subsubsection{Prior and posterior}\label{sec:Prior_and_posterior}
To minimize the risk of biased results, we assign flat, non-informative, priors to the parameters of the model.
This approach differs from the one of Ross17, \cite{Anderson:2013zyy}, \citet{XuPadmanabhan+2012}, which assigned a Gaussian prior to the shape parameter $B$. 
The downside of this choice is the risk of hitting a local maxima of the 
likelihood, $\mathcal{P}_\mathcal{L}(\mathbf{p})$, hence of increasing the uncertainty of the parameters estimate.
 To minimize this effect, we infer the median and the variance of the $i$-$th$ parameter $p_i$ by performing a robust-sigma estimation \citep[Section~3.2 in][]{Longobardi2015a} on the corresponding marginalized one-dimensional posterior 
\begin{equation}\label{eq:posterior}
    \mathcal{P}_\mathcal{L}(p_i)=\int_{D}dp_1 \dots dp_{i-1}dp_{p+1}\dots dp_\mathrm{N_p} \mathcal{P}_\mathcal{L}(\mathbf{p}),
\end{equation} 
where $D$ is the interval over which the prior of all $j$ parameters ($j\neq i$) are defined. We do so by defining the $2\sigma$-clipped posterior, $\mathcal{P}_\mathrm{2\sigma}(p_i)=\mathcal{P}_\mathcal{L}(p_i)\Theta\left(p_i-\langle p_i \rangle_\mathrm{f}-\sigma_{\mathrm{f},i}\right)\Theta\left(\langle p_i \rangle_\mathrm{f}-\sigma_{\mathrm{f},i}-p_i\right)$, where $\langle p_i \rangle_\mathrm{f}$ and $\sigma_{\mathrm{f},i}$ are the mean and variance of $\mathcal{P}_\mathcal{L}(p_i)$ prior to sigma-clipping, and $\Theta$ is the Heaviside step function. The variance of  $\mathcal{P}_\mathrm{2\sigma}(p_i)$ is assigned to $p_i$, scaled by a numerical factor determined from Monte Carlo simulations to complete the $2\sigma$-clipped distribution to a complete Gaussian.

\vspace{-0.2cm}
\section{\lowercase{e}FAM reconstruction in the mock catalogs. Validation tests}\label{sec:Results}

\begin{figure*}
\centering
\subfloat
{\includegraphics[width=1.\columnwidth]{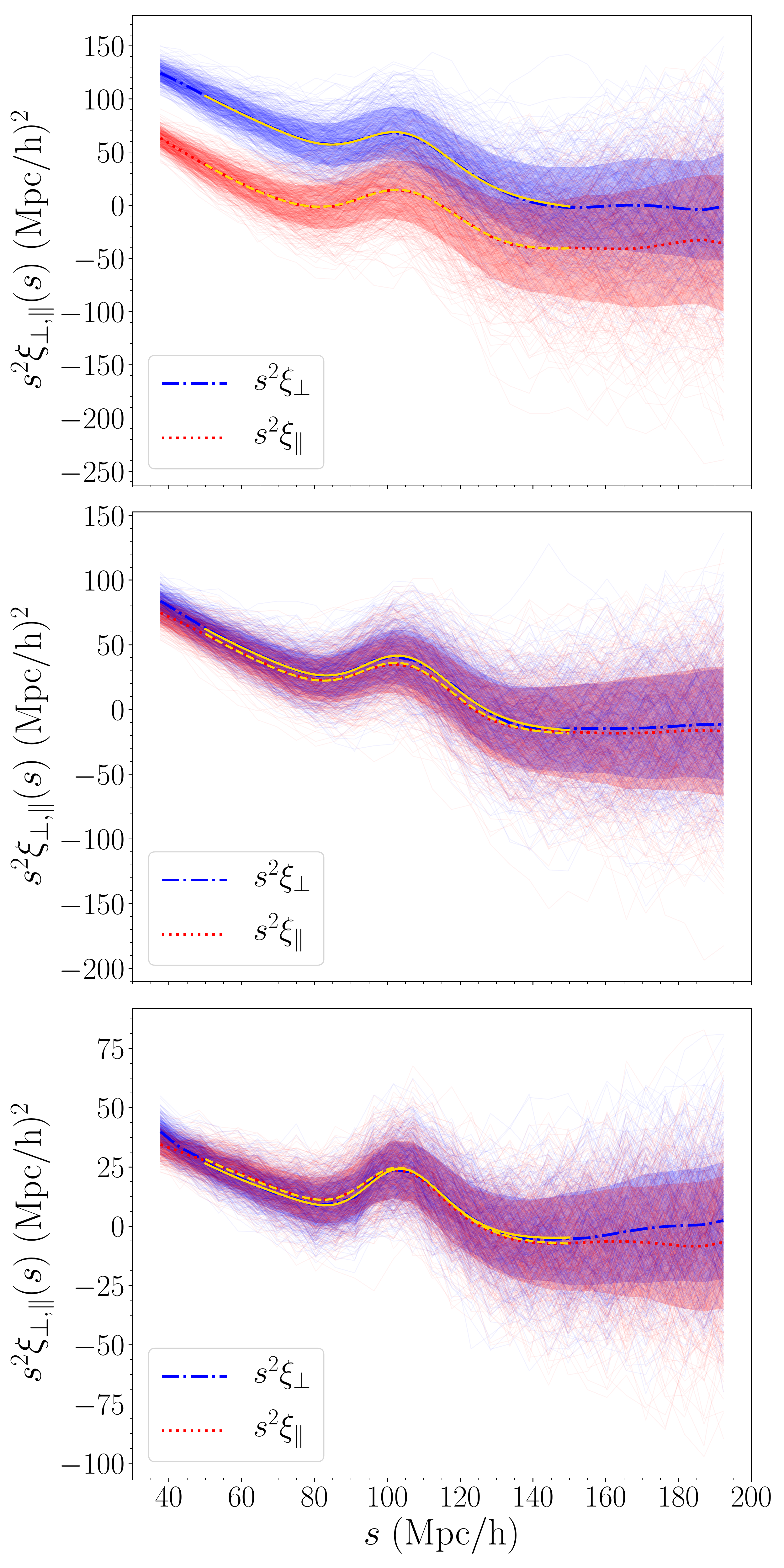}}
\subfloat
{\includegraphics[width=1.\columnwidth]{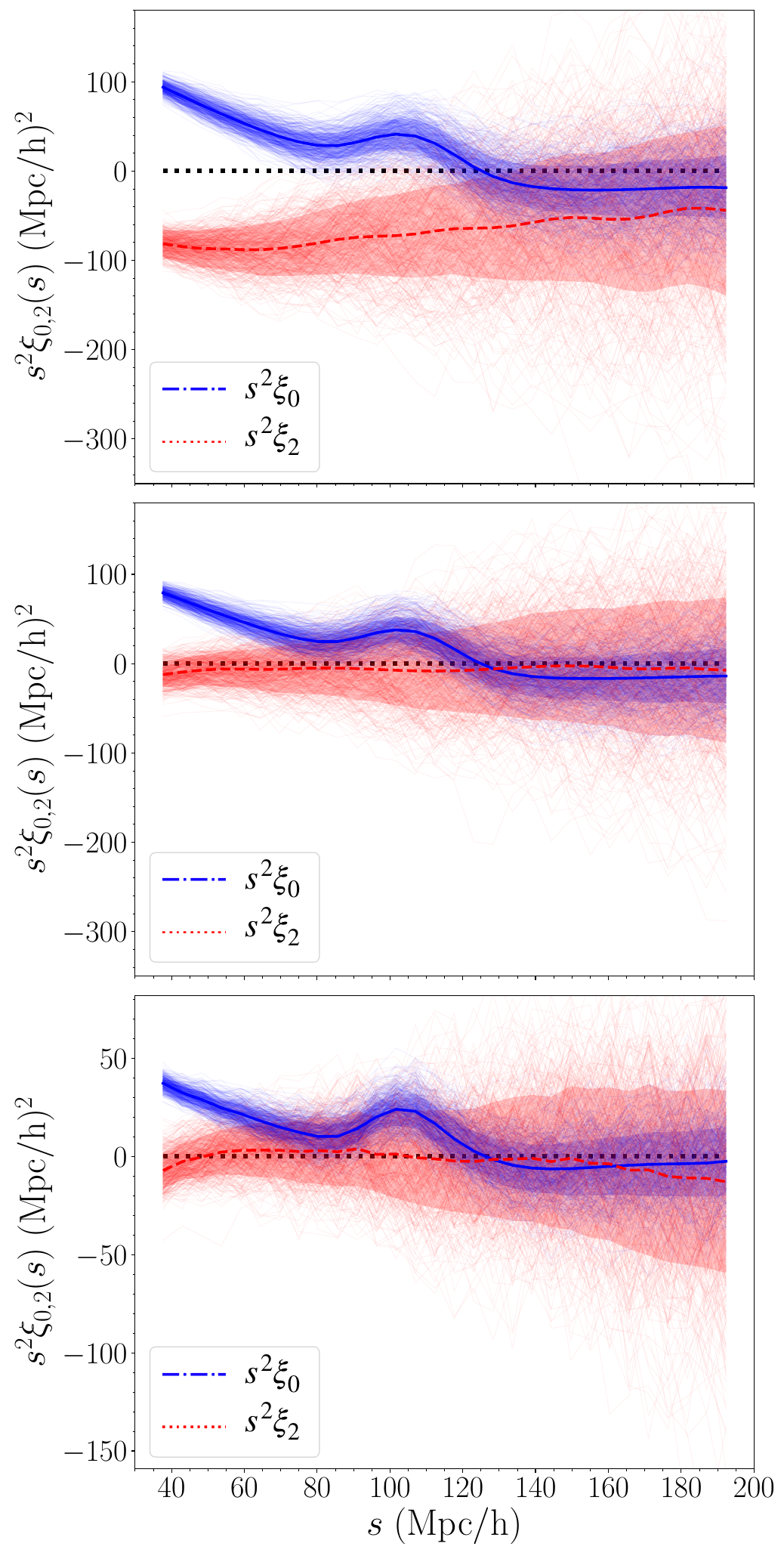}}
\caption{Two-point correlation function wedges (left panels) and multiples (right) estimated from 400 mock catalogues. \emph{Top}: mock galaxies are placed at their observed redshift positions, \emph{Middle}: galaxies are 
placed at the reconstructed real-space positions at the same 
redshift of the survey.
 \emph{Bottom}: galaxies are located at their back-in-time reconstructed real-space positions. Thin curves show the wedges measured in each mock. Thick curves with different line-styles show their average.
 Shaded regions show the 1$\sigma$ uncertainty strip. Yellow curves on the right hand panels are the best fitting models. 
 } 
\label{fig:MDPatchy_wedgesAmult}. 
\end{figure*}

In this section, we validate and investigate the performance of eFAM reconstruction focusing on two key aspects: its ability to remove RSD, and that of improving the signal-to-noise ratio of the BAO feature. For the following tests, we use the 400 MD-Patchy mocks mimicking the SDSS catalog described in Section~\ref{subsec:mocks}.

\vspace{-0.4cm}
\subsection{From redshift to real-space}\label{subsec:real_xi}
\begin{table*}
	\centering
	\caption{Best fit parameters and their 1$\sigma$ uncertainties estimated to the mean clustering wedges for the three cases considered: redshift-space sample (Obs), eFAM reconstructed real-space sample at the redshift of the survey (RecZ), and back-in-time reconstructed sample (RecL).}
	\label{tab:Velrec}
	\begin{tabular}{lccccccc}
		\hline
		sample & $\alpha_\perp$ & $\alpha_\parallel$ & $\Sigma_\mathrm{\perp}(h^{-1}\mathrm{Mpc})$ & $\Sigma_\mathrm{\parallel}(h^{-1}\mathrm{Mpc})$ & $f$ & $\Sigma_{\mathrm{s}}(h^{-1}\mathrm{Mpc})$ \\
		\hline
		
		Obs &0.997  $\pm$  0.003 &1.011 $\pm$  0.007 &7.37 $\pm$  0.45 & 10.20 $\pm$ 1.82 & 0.51 $\pm$ 0.13 &4.43$\pm$ 1.83 \\
		
		RecZ & 0.997 $\pm$0.003 &1.011$\pm$ 0.006 &7.79 $\pm$ 0.36 & 9.12$\pm$ 0.68 &0.07 $\pm$ 0.06 &1.09$\pm$1.18\\
		
	    RecL  & 0.997 $\pm$ 0.002 &1.010 $\pm$ 0.004 & 5.21 $\pm$ 0.39 & 6.93 $\pm$ 0.53 &0.06 $\pm$ 0.05 &1.02 $\pm$ 0.88\\
	    \bottomrule
	 \end{tabular}
\end{table*}

To assess the quality of the redshift-to-real space reconstruction, we apply eFAM to each of the 400 mock catalogues and measure the 2PCF monopole and quadrupole moments as well as the clustering wedges of the observed (Obs) and reconstructed (RecZ and RecL) samples defined in  Section~\ref{subsec:estimator}.
To minimize the effect of tidal fields, we discard all objects within $d_\mathrm{Buffer}=200h^{-1}\mathrm{Mpc}$ from the edges of the survey.
The results are illustrated in Figure~\ref{fig:MDPatchy_wedgesAmult}. 
Top panels illustrates the results obtained from the Obs sample. Middle and Bottom panel refer to the RecZ and RecL cases, respectively.
Panels on the left show the 2PCF parallel (red) and perpendicular (blue) clustering wedges. Panels on the right illustrate the 2PCF monopole and the quadrupole moments. All quantities are multiplied by the square of the spatial separation $s$. 
Thin curves show the measurement for each of the 400 mocks, thick curves show their average values, and shaded areas indicate the standard deviation among the mocks, corresponding to the square root of the diagonal elements of the covariance matrix.
Finally, yellow thick curves on the left panels represent the best-fit model for the clustering wedges.

In redshift-space (Obs, top-left), the parallel and perpendicular wedges are significantly different from each other. This discrepancy is expected. Indeed, RSD simultaneously reduce the amplitude of the clustering signal along the line-of-sight and boost up the amplitude of the perpendicular wedge. Moreover, RSD generate a quadrupole moment that is clearly visible in the top-right panel.
A visual inspection of the mid panels (RecZ) shows the success of eFAM at removing RSD; the two wedges are now in agreement, almost superimposed to each other (left panel), and the quadrupole moment is consistent with zero (marked by the black, dotted horizontal line in the right panels).
Remarkably enough, the statistical isotropy is restored down to the smallest scales shown in the plot, $\sim 40h^{-1}\mathrm{Mpc}$, demonstrating that eFAM is indeed able to model dynamics well into the non-linear regime. 

\begin{figure}
\centering
\includegraphics[width=1\columnwidth]{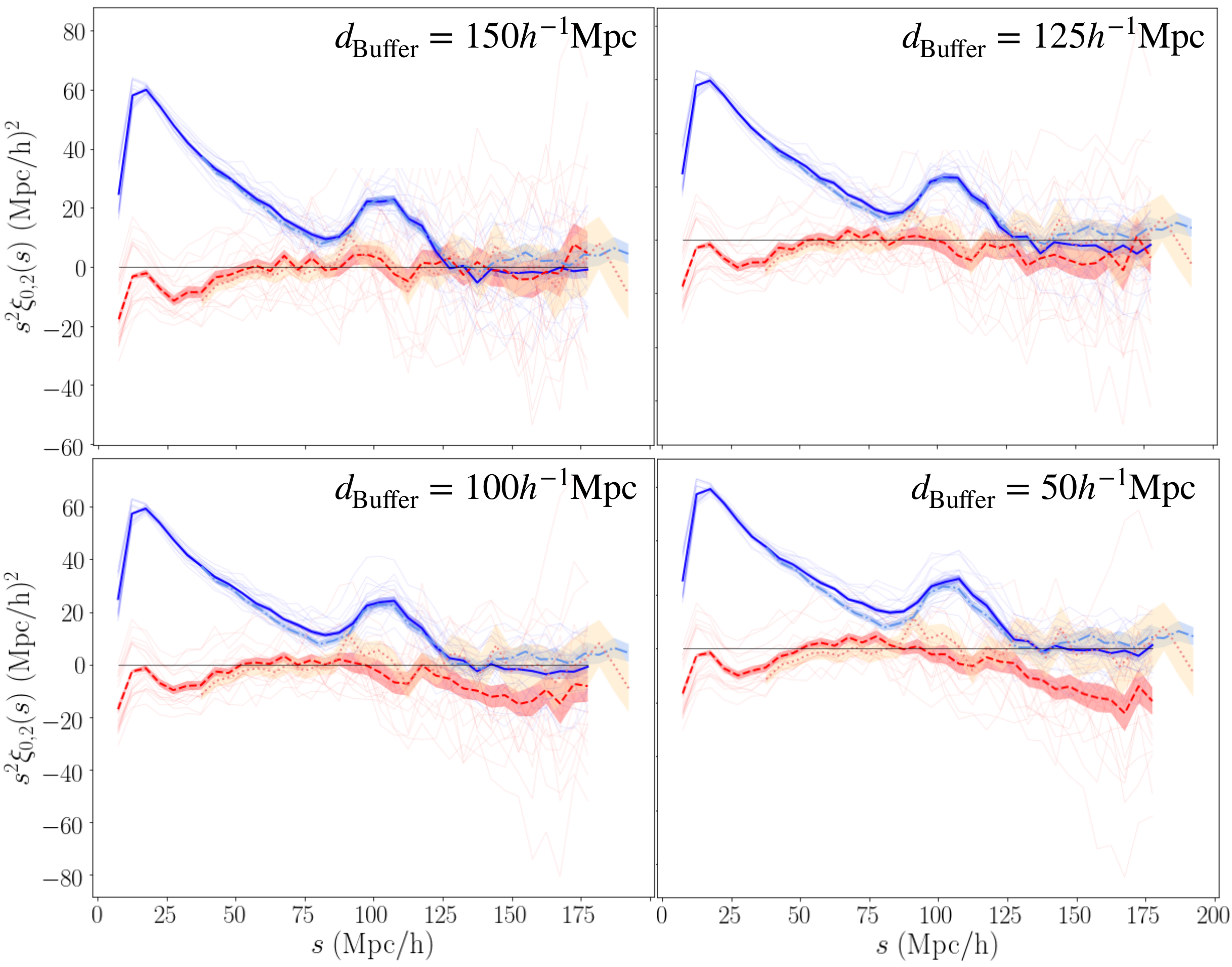}
\caption{Effect of the external (super-survey) tidal field on reconstruction. Mean monopole (blue-solid lines) and quadrupole (red-dashed lines) of the 2PCF averaged over 20 mocks measured for different choices of the discarded buffer region, $d_\mathrm{Buffer}$. Light-blue dot-dashed lines and orange-dotted lines mark the mean monopole and quadrupole as estimated for $d_\mathrm{Buffer}=200h^{-1}\mathrm{Mpc}$. }\label{fig:Buffer}
\end{figure}
A successful reconstruction must remove RSD at all epochs, not only at the redshift of the survey. We check for this behaviour estimating the isotropic 2PCF of the RecL sample at a conveniently high redshift. Here, we stop the reconstruction at  $z\sim 40$ corresponding to the highest redshift for which the measured $\Sigma_\parallel$ and $\Sigma_\perp$ have the smallest, non-negative value (see Section~4 of \cite{Sarpa:2018ucb} for a detailed discussion on the definition and justification of the maximum redshift of the eFAM reconstruction). 
The 2PCF wedges and multipoles of the RecL sample are plotted in the bottom panels of Figure~\ref{fig:MDPatchy_wedgesAmult}. Similarly to the RecZ case, RSD are successfully removed, although some residual anisotropy is seen at very large separations. We interpret this feature as a signature of imperfect correction for external tidal fields. To check this hypothesis, we repeated the test varying the depth of the buffer region $d_\mathrm{Buffer}$. The results are shown in Figure \ref{fig:Buffer}.
Thick lines represent the average monopole (blue-solid) and quadrupole 
(red-dashed) moments of the 2PCF measure in 20 independent mock catalogues. Shaded areas represent the standard deviation. The value of $d_\mathrm{Buffer}$ used in each reconstruction is indicated in each panel.
For all the values of $d_\mathrm{Buffer}$ we testes, we also show the reference case $d_\mathrm{Buffer}=200 \ h^{-1}$ Mpc (light blue and light red curves and areas for the monopole and quadrupole moments, respectively).
The quality of the reconstruction improves when using larger $d_\mathrm{Buffer}$, i.e. when discarding an increasing fraction of objects in the sample. The main effect of the spurious tidal field is a deceptive quadruple at large separation that is particularly significant for $d_\mathrm{Buffer}=50 \ h^{-1}$ Mpc, and progressively vanishes when moving towards larger sizes of the buffer region. A second effect is a tiny artificial correlation at the deep's scale in the monopole.
Based on this test, we decide to set $d_\mathrm{Buffer}=200 \ h^{-1}$ Mpc being confident that with this choice RSD are effectively removed down to  $\sim 40 \ h^{-1}$ Mpc and the measured monopole signal is robust to less conservative choices of $d_\mathrm{Buffer}$.

A more quantitative assessment of the quality of the RSD removal is provided by the likelihood analysis. Table~\ref{tab:Velrec} shows the values of the best fitting parameters to the mean clustering wedges estimated in the 400 mocks and their 1$\sigma$ uncertainty.
The parameters that quantify RSD are the growth rate $f$ and the velocity dispersion $\Sigma_S$. Their value can be estimated by measuring the wedges in redshift-space (Obs sample, first row in the Table).
The efficient removal of RSD by eFAM is testified by the 1$\sigma$ compatibility of both parameters with zero the RecZ and RecL samples alike. To perform the likelihood analysis we have considered the range of separation $s=[50,150]h{-1}$Mpc and the best fitting curves are drawn in yellow color in Figure \ref{fig:MDPatchy_wedgesAmult}.

\begin{figure}
\centering
\includegraphics[width=.9\columnwidth]{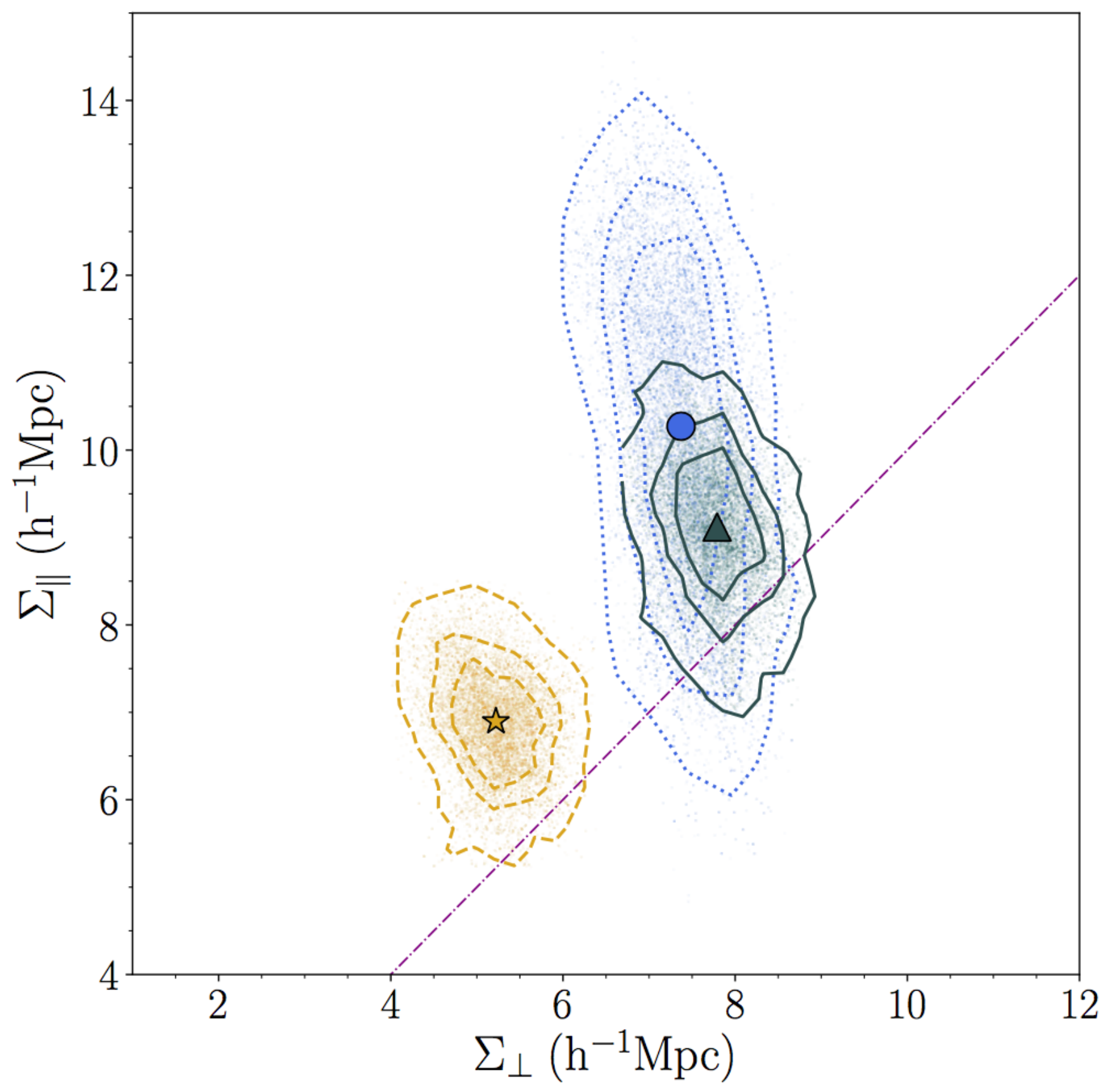}
\caption{Marginalized probability contours of $\Sigma_\perp$, $\Sigma_\parallel$ estimated from the fit of the mean clustering wedges of the mocks obtained from the Obs sample (blue dotted contours), the RecZ sample (dark green solid contours) and the RecL sample (gold, dashed contours). The large symbols (circle, triangle and star) show the value of the best fitting parameters for, respectively, the Obs, RecZ and RecL cases. The contours are drawn at 1, 2, and 3$\sigma$ confidence levels. Small dots show the values obtained at each step of the Markov-Chain. The purple, dot-dashed line is drawn for reference to indicate $\Sigma_\parallel=\Sigma_\perp$.
}\label{fig:Snl}
\end{figure}

As shown by \cite{Padmanabhan+2012}, another test of the quality of the reconstruction is to compare the relative magnitude of the two damping parameters, $\Sigma_{\perp,\parallel}$, before and after reconstruction. Indeed, to move galaxies to their correct real-space position one should remove all sources of anisotropy, not just the RSD. As result, a successful reconstruction should bring the ratio $\Sigma_\parallel/\Sigma_\perp$ closer to unity \citep{Snl_rsd2018JCAP...07..053I}. Since eFAM is designed to account for non-linear motions, we also expect that the absolute magnitude of both parameters should be reduced. The contour levels of the marginalized, joint probability distribution of $\Sigma_\parallel$ and $\Sigma_\perp$ shown in Figure~\ref{fig:Snl} prove that this is indeed the case. The blue-dotted contours show the results obtained before reconstruction for the Obs sample while the large blue-dot show the median values. The magnitude of the two parameters is rather large and their ratio is about 2$\sigma$ apart from unity (dot-dashed purple line). Dark-green continuous curves (and the green-triangle) show the results after the eFAM reconstruction for the RecZ sample, i.e. after removing RSD but with non-linear effects still present. The magnitude of the two parameters remains the same but their ratio is closer to unity. Gold-dashed contours show the RecL case, after eFAM back-in-time reconstruction. In this case, also the magnitude of both parameters is significantly reduced.
Besides, we notice that the reconstruction significantly diminishes the uncertainty in the measured $\Sigma_\parallel$. This improvement indicates that eFAM efficiently lifts the degeneracy between $\Sigma_\parallel$ and the $A_p(s)$ parameters describing the broad-band shape of the 2PCF while restoring statistical isotropy.

The validation tests performed in this section are designed to quantify the ability of eFAM to remove the anisotropies generated by linear and non-linear motions rather than the ones arising from an incorrect choice of the fiducial cosmology. For this reason, we assume the same cosmological model as the mock catalogues. To test for the absence of Alcock-Paczynski distortions, we do not set the dilation parameters equal to unity but treat instead $\alpha_\parallel$ and $\alpha_\perp$ as free parameters in the fit and check if their best-fit value is indeed consistent with unity. 
The results listed in Table~\ref{tab:MDPatchy_alpha} show that this is indeed the case. The largest departures are at $\sim 2\sigma$ level for $\alpha_\parallel$. 
These results prove that eFAM successfully removes RSD from the 2PCF and allow us to use the real-space template ($f=\Sigma_s=0$) to model the reconstructed 2PCF of the data in Section~\ref{sec:ResultsData}.

\vspace{-0.2cm}
\subsection{Acoustic scale measurements}\label{subsec:AcoustiMocks}

\subsubsection{Alcock-Paczynski distortions}
The second part of the validation process focuses on the ability of eFAM to estimate the dilation parameters $\alpha_\parallel$ and $\alpha_\perp$. For this purpose, we perform a likelihood analysis in each mock using the clustering wedges measured in the separation range $s \in [50,150] h^{-1}\mathrm{Mpc}$. In this analysis, the values of $f$, $\Sigma_\parallel$, and $\Sigma_\perp$ are kept constant. For the Obs sample they are set equal to their best-fit values listed in Table~\ref{tab:Velrec}, while for RecZ and RecL we set them equal to zero. To reduce the uncertainty in the estimated $\alpha_\perp$ and $\alpha_\parallel$, we searched for the largest volume of the sample, namely for the minimum value of $d_\mathrm{Buffer}$, yielding an unbiased estimation of the real-space density field.  We find $d_\mathrm{Buffer}=125h^{-1}$Mpc to be a good compromise between sample volume size and systematic errors driven by edge effects. A smaller $d_\mathrm{Buffer}$ would not bias the position of the BAO peak, albeit it would reduce the accuracy of the RSD modelling for $s> 150h^{-1}$Mpc, as shown in the top-right panel of Figure~\ref{fig:Buffer}. Hereafter, we will only consider the Obs and RecL samples.

\begin{figure}
\centering
\includegraphics[width=.8\columnwidth]{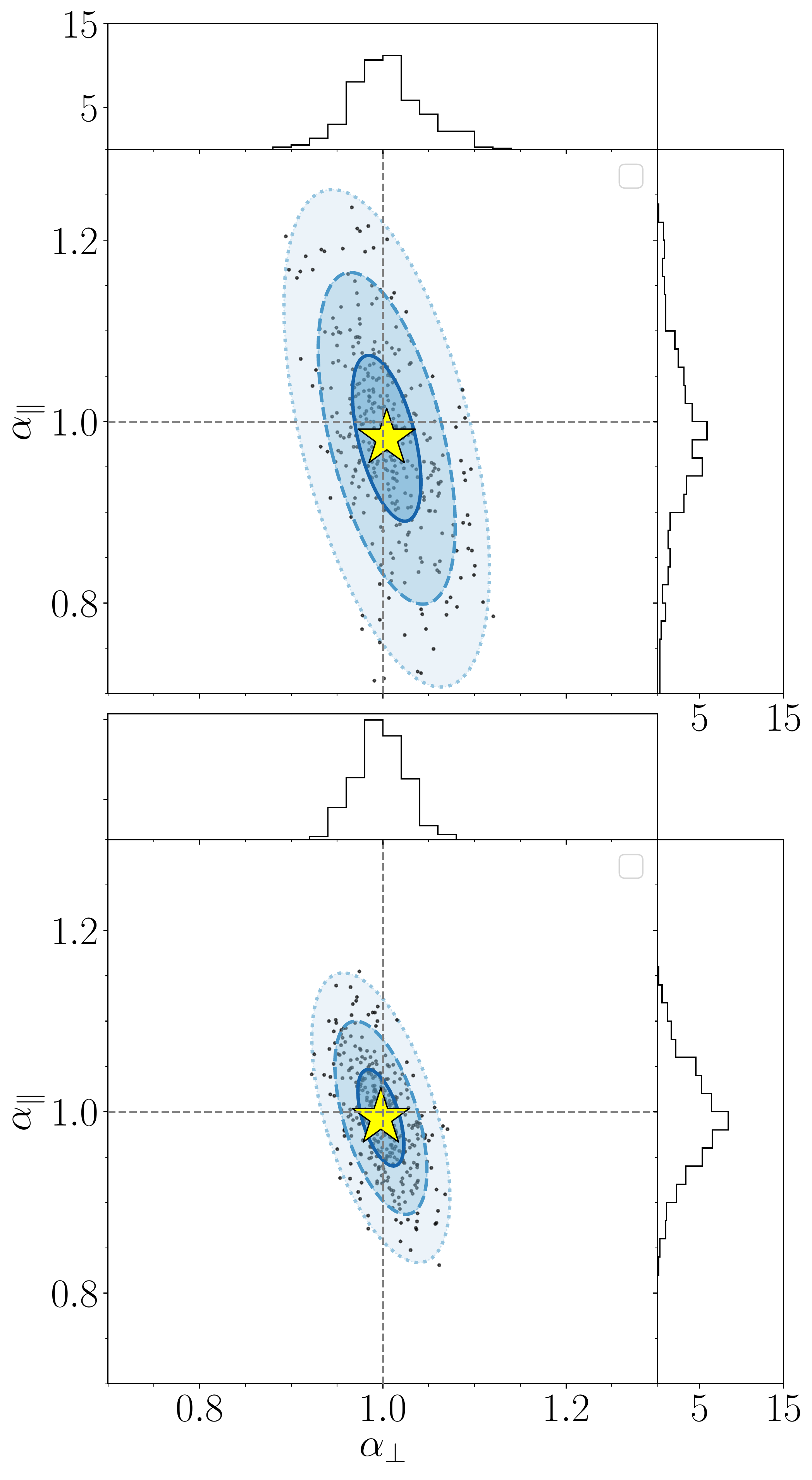}
\caption{Best fit $\alpha_\parallel$ an $\alpha_\perp$ values estimated from the clustering wedges in the range $s=[50,150]h^{-1}$Mpc measured in in each MD-Patchy mocks (dots)
for the Obs (Top) and RecL (Bottom) cases. The histograms on the top and on the right part of the panels show the one-dimensional distribution functions of $\alpha_\perp$ and $\alpha_\parallel$, respectively.
Elliptical contours and blue-shaded areas are drawn at the 68, 95, and 99 per cent confidence level of the best fitting Gaussian bivariate, centered at the starred symbol.Grey-dashed lines show the expected $\alpha_\perp=1$ and $\alpha_\parallel=1$ values.}
\label{fig:MDPatchy_alpha_alpha}
\end{figure}
\begin{figure}
\centering
\includegraphics[width=.8\columnwidth]{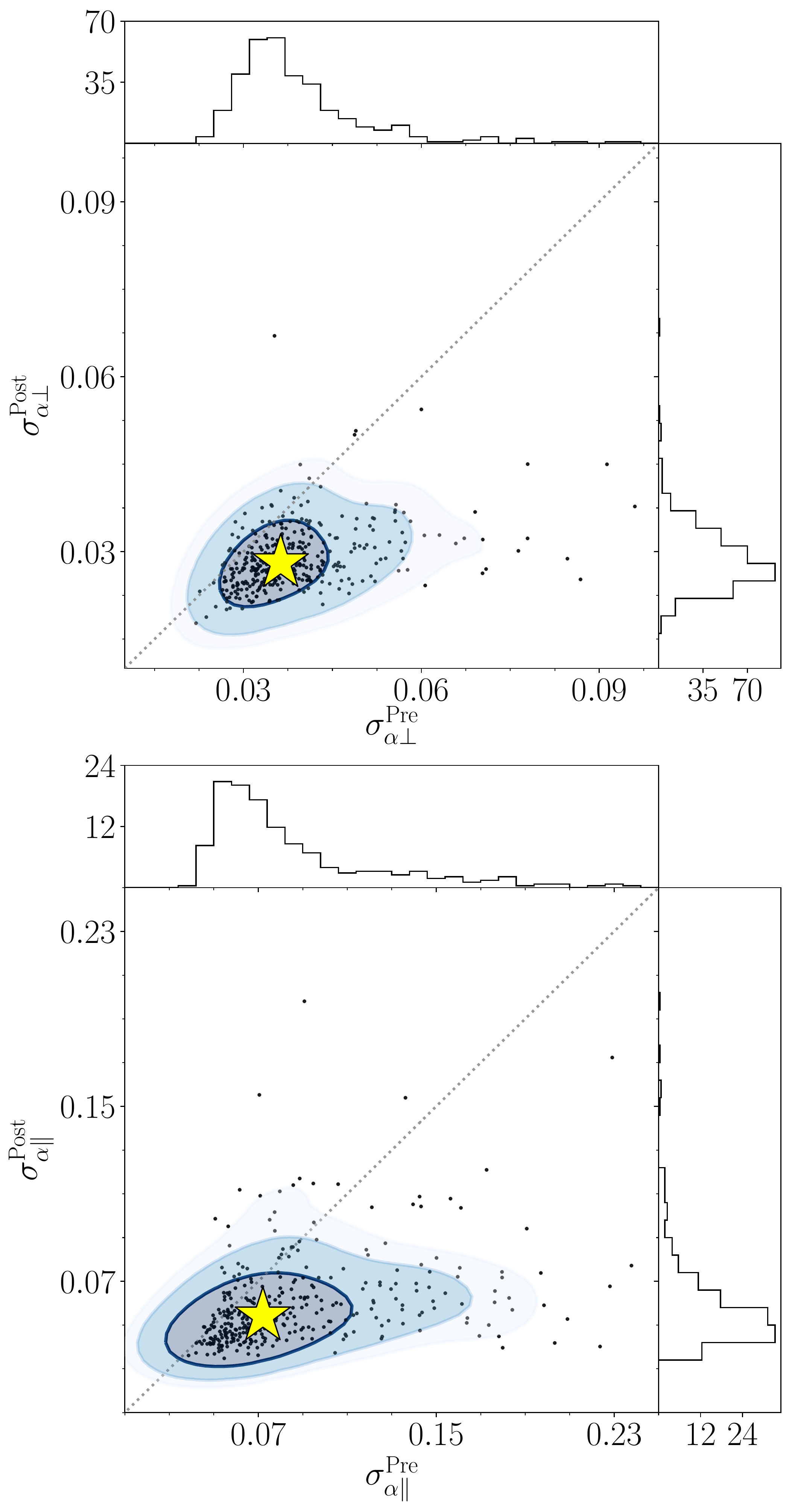}
\caption{Error on the perpendicular (top) and parallel (bottom) dilation parameters,
$\sigma_{\alpha_\perp}$ (top panel) and $\sigma_{\alpha_\parallel}$, 
before and after eFAM-reconstruction estimated from the clustering wedges. 
Each dot shows the results from each mock catalog. Shaded areas contain 68, 95, and 99 of the points. The average  value of the distribution is represented by a starred symbol.
Histograms depict the 1D marginalized distribution of both errors, before (histogram on top) and after (histogram on the side) eFAM reconstruction. Dotted lines show the case in which reconstruction gives no improvement.}
\label{fig:MDPatchy_Salpha_Salpha}
\end{figure}

The results of our reconstruction of the acoustic feature are shown in Figure~\ref{fig:MDPatchy_alpha_alpha} 
and summarized in Table~\ref{tab:alpha_alpha}. Each point in Figure~\ref{fig:MDPatchy_alpha_alpha} represents the best-fit value obtained from each mock catalog before (top) and after (bottom) eFAM reconstruction. 
In both cases, we fit the distribution of points in the  $(\alpha_\perp,\alpha_\parallel)$ plane with a bi-variate Gaussian for which we show its maximum (star symbol) and the 68, 95 and 99 per cent confidence levels.
The histograms on the top and right side of each panel show the marginalized probability distribution function of $\alpha_\perp$ and $\alpha_\parallel$, respectively.
Since our fiducial cosmology is set to the one of the mocks, the expected values are $\alpha_\perp = \alpha_\parallel =1$.

The distribution of the best fit $\alpha_\parallel $ and $\alpha_\perp $ in the Obs sample (top panel) is $\gtrsim 30$ per cent broader than in the RecZ sample (bottom panel). Alongside, the central value of the bi-variate Gaussian is more biased pre-reconstruction, $\sim 1.7$ per cent along the $\alpha_\parallel$ direction, than post-reconstruction; see Table~\ref{tab:alpha_alpha}. 
Not expecting any Alcock-Paczynski distortion, we interpret both the bias and the strong uncertainty in the pre-reconstruction estimates of $\alpha_\parallel$ and  $\alpha_\perp$ as a sign of the inadequacy of the fiducial RSD model (Equation~\ref{eq:pk_rsd}) to account for non-linear effects. Remarkably enough, the fact that our reconstruction significantly decreases both the dispersion and the offset of the dilation parameters indicates that eFAM successfully accounts for both linear and non-linear motions in the range of scales considered in the analysis.
The residual bias and dispersion affecting in the post-reconstruction measurements are probably a consequence of the imperfect modelling of the external mass distribution and its tidal field.

\begin{figure*}
\centering
\includegraphics[width=0.9\textwidth]{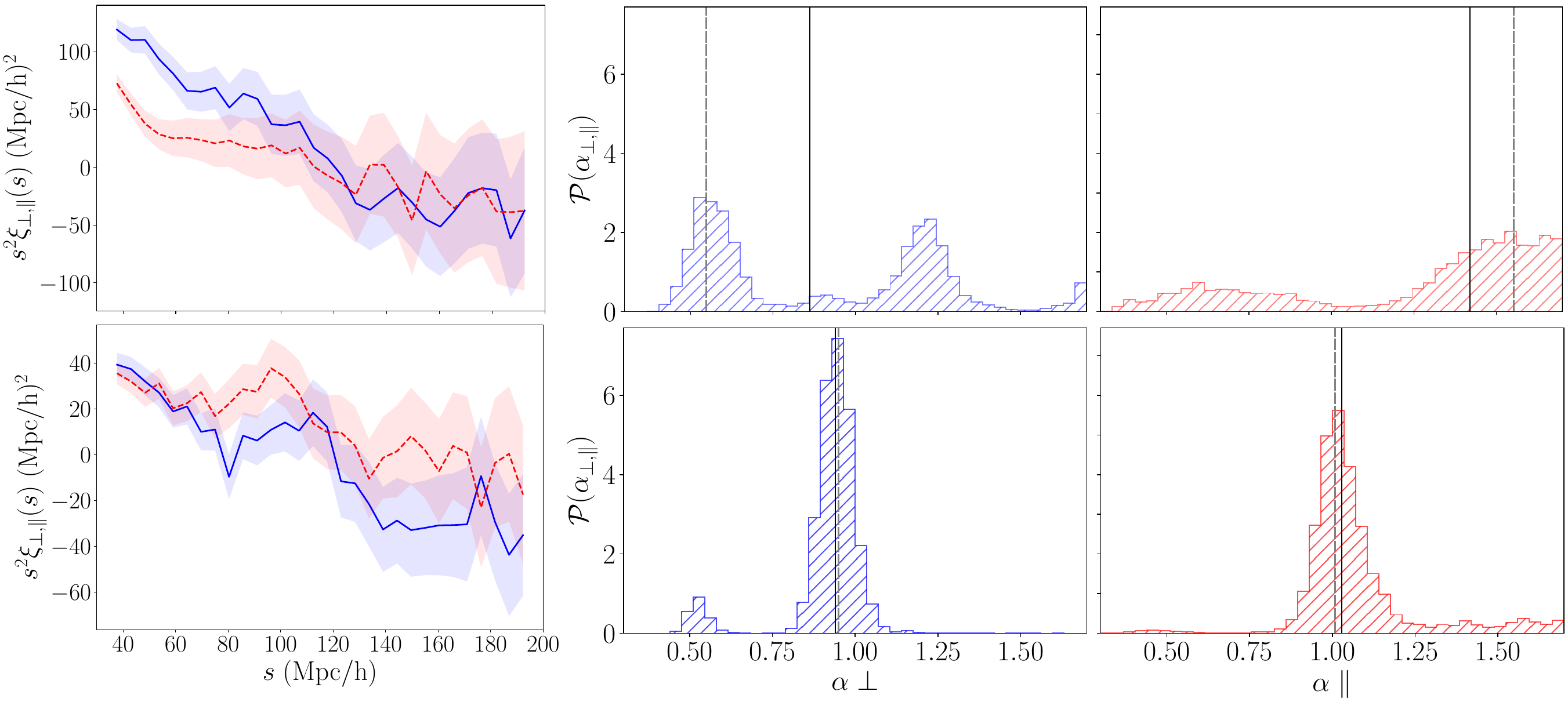}
\caption{Modelling the clustering wedges in a single mock realization.
\emph{Left panels:} parallel (red-dashed lines) and perpendicular (blue-solid lines) clustering wedges multiplied by the square of the spatial separation $s$.
Shaded areas represent the standard deviation.
\emph{Central and Right panels:} marginalized 1-dimensional probability distribution of $\mathcal{P}_{\alpha_\perp}$ and $\mathcal{P}_{\alpha_\parallel}$, respectively.  \emph{Top panels:} Results obtained from the Obs sample.
  \emph{Top panels:} Results obtained from the RecL sample.
Solid-black vertical lines mark the median values of the 1-dimentional $\mathcal{P}$ while dashed-gray lines demote the best fit values.}

\label{fig:Single_mock}
\end{figure*}

To further investigate the sources of uncertainty we have considered the results of the likelihood analysis performed on a single mock in which no clear BAO peak is detected in the clustering wedges before reconstruction (top right panel of Figure~\ref{fig:Single_mock}).
The histograms in the central and right top panels show the one-dimensional posterior probability function for $\alpha_\perp$ and $\alpha_\parallel$, respectively, obtained from the Markov Chain procedure. With no BAO feature detected, neither distribution exhibit a well defined, main peak. Instead, $\mathcal{P}_{\alpha_\perp}$ (left) has two maxima of comparable amplitude whereas $\mathcal{P}_{\alpha_\parallel}$ shows a broad peak extending beyond the range allowed for this parameter. As a result, one obtains a noisy and biased estimate of both parameters.
The situation dramatically improves after eFAM reconstruction (bottom panels). 
A BAO signature is now visible in both wedges (bottom right) and, consequently, the posterior probability distributions of both dilation parameters show a well defined, sharp maximum centered on the expected values $\alpha_\perp=\alpha_\parallel=1$.

\begin{table*}
	\centering
	\caption{Best fit vs. separation range. Median values and variance of $\alpha_\perp$, $\alpha_\parallel$, $\sigma_{\alpha_\perp}$, and $\sigma_{\alpha_\parallel}$ estimated from clustering wedges in different separation ranges.}\label{tab:alpha_alpha}
	\label{tab:MDPatchy_alpha}
	\begin{tabular}{llcccccccc}

		\midrule
		fitting range &sample & $\langle\alpha_\perp\rangle-1$ & $S_{\langle\alpha_\perp\rangle}$& $\langle\sigma_{\alpha_\perp}\rangle$&$S_{\langle\sigma_{\alpha_\perp}\rangle}$&$\langle\alpha_\parallel\rangle-1$ & $S_{\langle\alpha_\parallel}\rangle$& $\langle\sigma_{\alpha_\parallel}\rangle$ &$S_{\langle\sigma_{\alpha_\parallel}\rangle}$\\
		\hline
		$[50,150]h^{-1}$Mpc &Obs &0.002 & 0.036 &0.031 &\begin{tabular}{@{}c@{}}+0.017\\  -0.004\end{tabular}& -0.017 &0.081& 0.053 &\begin{tabular}{@{}c@{}}+0.067\\  -0.005\end{tabular}\\
		
		&RecL &-0.002&0.024& 0.024& \begin{tabular}{@{}c@{}}+0.013\\  -0.002\end{tabular}& -0.010 & 0.056&  0.049 &\begin{tabular}{@{}c@{}}+0.032\\  -0.006\end{tabular}\\
		\hline
		$[40,150]h^{-1}$Mpc &Obs &0.002&  0.035 &0.032 &\begin{tabular}{@{}c@{}}+0.024\\  -0.004\end{tabular}& -0.026& 0.088& 0.054 &\begin{tabular}{@{}c@{}}+0.061\\  -0.004\end{tabular} \\
		
		&RecL &-0.002& 0.025& 0.027& \begin{tabular}{@{}c@{}}+0.007\\  -0.004\end{tabular} & -0.011 &0.051&  0.043 &\begin{tabular}{@{}c@{}}+0.030\\  -0.004\end{tabular} \\
		\hline
	
		$[25,150]h^{-1}$Mpc &Obs &0.004 & 0.036 &0.033 &\begin{tabular}{@{}c@{}}+0.017\\  -0.004\end{tabular}& -0.032 &0.091& 0.059 &\begin{tabular}{@{}c@{}}+0.065\\ 
		-0.007\end{tabular} \\
		
		&RecL &0.000& 0.023& 0.026& \begin{tabular}{@{}c@{}}+0.013\\  -0.003\end{tabular} & -0.010 &0.053&  0.048 &\begin{tabular}{@{}c@{}}+0.030\\  -0.004\end{tabular} \\
	\bottomrule

	\end{tabular}
\end{table*}

An alternative way to assess the impact of eFAM in reducing the uncertainty of $\alpha_\perp$ and $\alpha_\perp$ is that of considering the errors of these parameters estimated from the 
clustering wedges measured in a single mock catalog. 
The scatter-plot in Figure~\ref{fig:MDPatchy_Salpha_Salpha} compares the single-mock uncertainties $\sigma_{\alpha_\perp}$ (top panel) and $\sigma_{\alpha_\parallel}$ (bottom panel) estimated before the reconstruction (X-axis) to those estimated after the reconstruction (Y-axis). The histograms on top and on the side show the marginalized one-dimensional distributions of both quantities. Before the reconstruction, both distributions are skewed toward large errors. After reconstruction eFAM succeeds in reducing the
the skewness by $\sim 30$ per cent.

\subsubsection{Robustness to non-linear effects.}

 To test the robustness of eFAM reconstruction to the inclusion of small scales characterized by non-linear effects, we push the likelihood analysis performed in 
 the previous section down to separations as small as $s_{\rm min}=25h^{-1}$Mpc.
 The results of this test are summarized in Table~\ref{tab:alpha_alpha} where we list the best fit values of $\alpha_\perp$ and $\alpha_\parallel$ and their error for 
 $s_{\rm min}=25$ and $ 40h^{-1}$Mpc along with the reference case $s_{\rm min}=50h^{-1}$Mpc.
 We notice that the bias on $\alpha_\parallel$ measured before the reconstruction increases when pushing the correlation analysis down to smaller scales. This is not surprising and confirms the increasing inadequacy of the RSD model in accounting for the effect of non-linear peculiar velocities. The eFAM reconstruction reverse this trend. The 
 systematic error on $\alpha_\parallel$ remains the same, irrespective of the $s_{\rm min}$
considered. 
This result demonstrates the success of eFAM reconstruction and corroborates our conclusion that systematic errors on $\alpha_\parallel$ 
originate from the external tidal field rater than non-linear effects.

\vspace{-0.2cm}
\section{Reconstructing SDSS data}\label{sec:ResultsData}

\begin{figure}
\centering
\includegraphics[width=1\columnwidth]{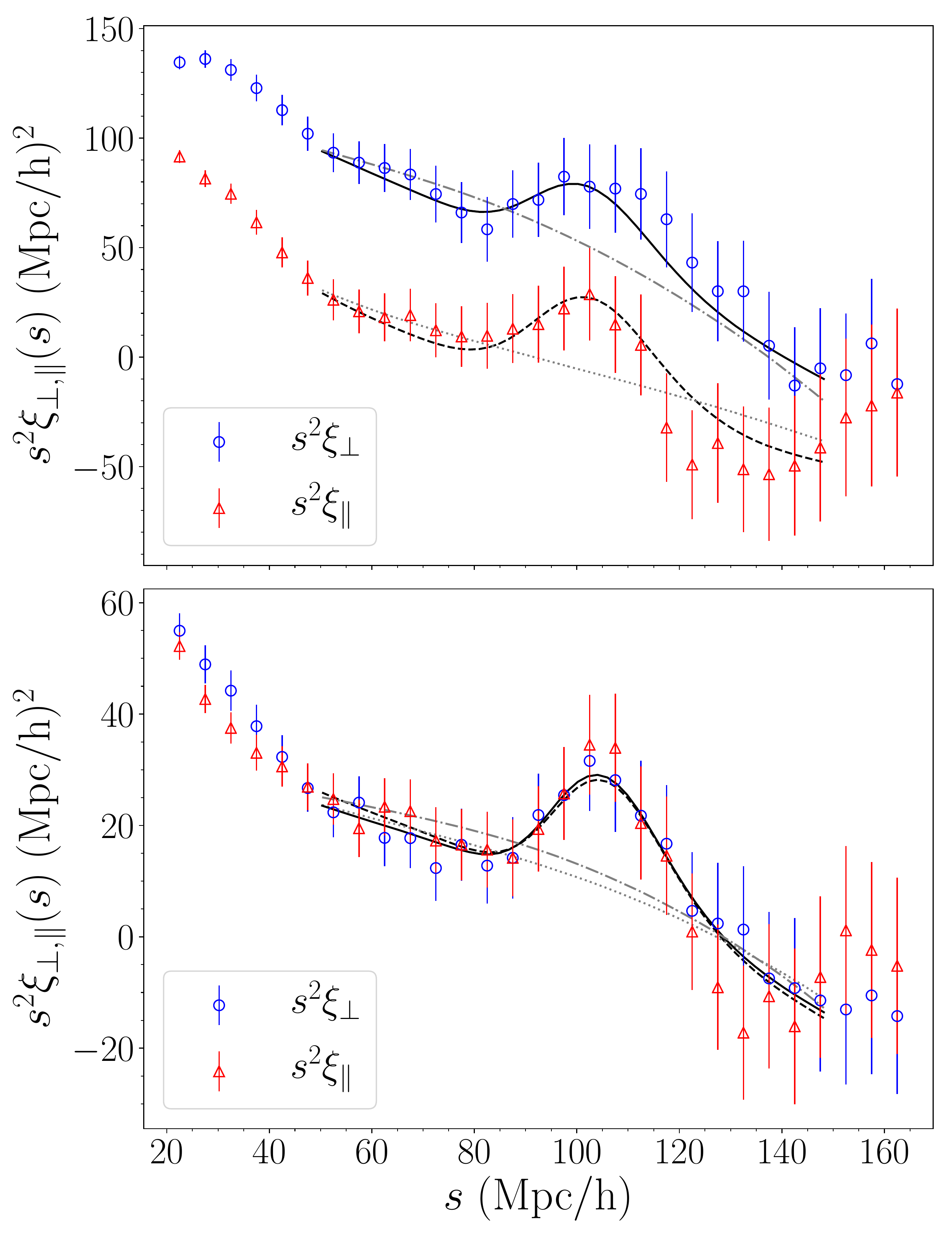}
\caption{Clustering wedges of the BOSS-DR12 Combined Sample before (top-panel), and after eFAM reconstruction (bottom-panel). 
Open blue circles show the measured perpendicular wedge. Open red triangles show the parallel wedge. The error-bars are the square root of diagonal of the covariance matrix estimated from the mocks.
Best fit clustering wedges models are also shown with black continuous curves.
Solid and dashed curves show the best fit model that include the BAO feature.  Dot and dot-dashed curves are best fitting models with no acoustic peak.}\label{fig:Data_wedges}
\end{figure}
\begin{figure}
\centering
\includegraphics[width=.8\columnwidth]{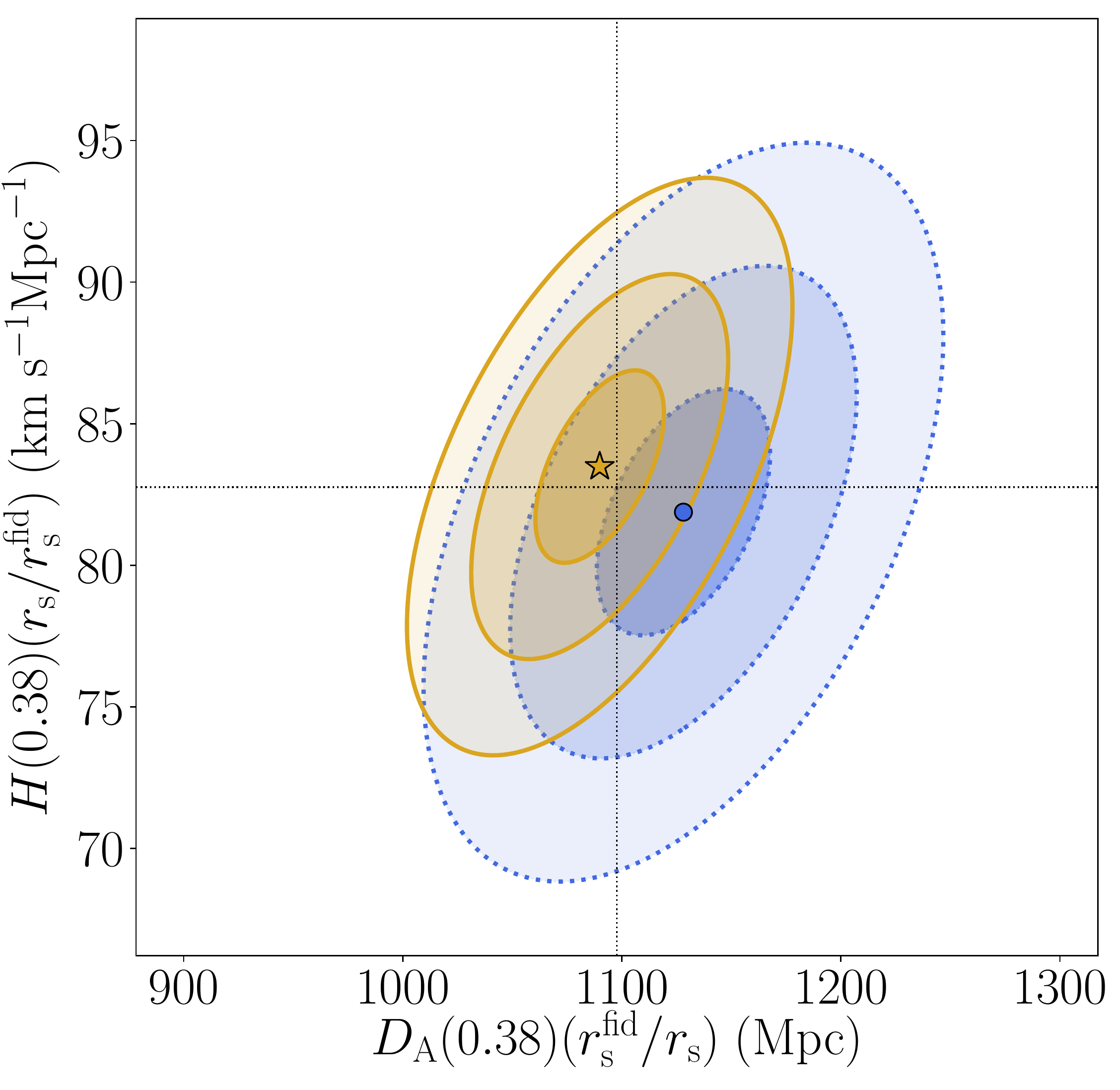}
\caption{marginalized likelihood contours for $D_\mathrm{A}r^\mathrm{fid}_\mathrm{s}/r_\mathrm{s})$ and $H (r_\mathrm{s}/r^\mathrm{fid}_\mathrm{s})$ from the clustering wedges analysis of the 
 SDSS-DR12 Combined Sample in the range $0.2<z<0.55$. Parameters are estimated at the effective redshift $z=0.38$.
 Contours and shaded areas show 1, 2, and 3$\sigma$ probability contours of a best fitting Gaussian bi-variate. 
 Blue contours: results before eFAM reconstruction. Yellow contours: results after eFAM reconstruction. Large blue and star symbols show the best fit values for the two cases. }\label{fig:Data_cont}
\end{figure}
\begin{figure*}
\centering
\includegraphics[width=.9\textwidth]{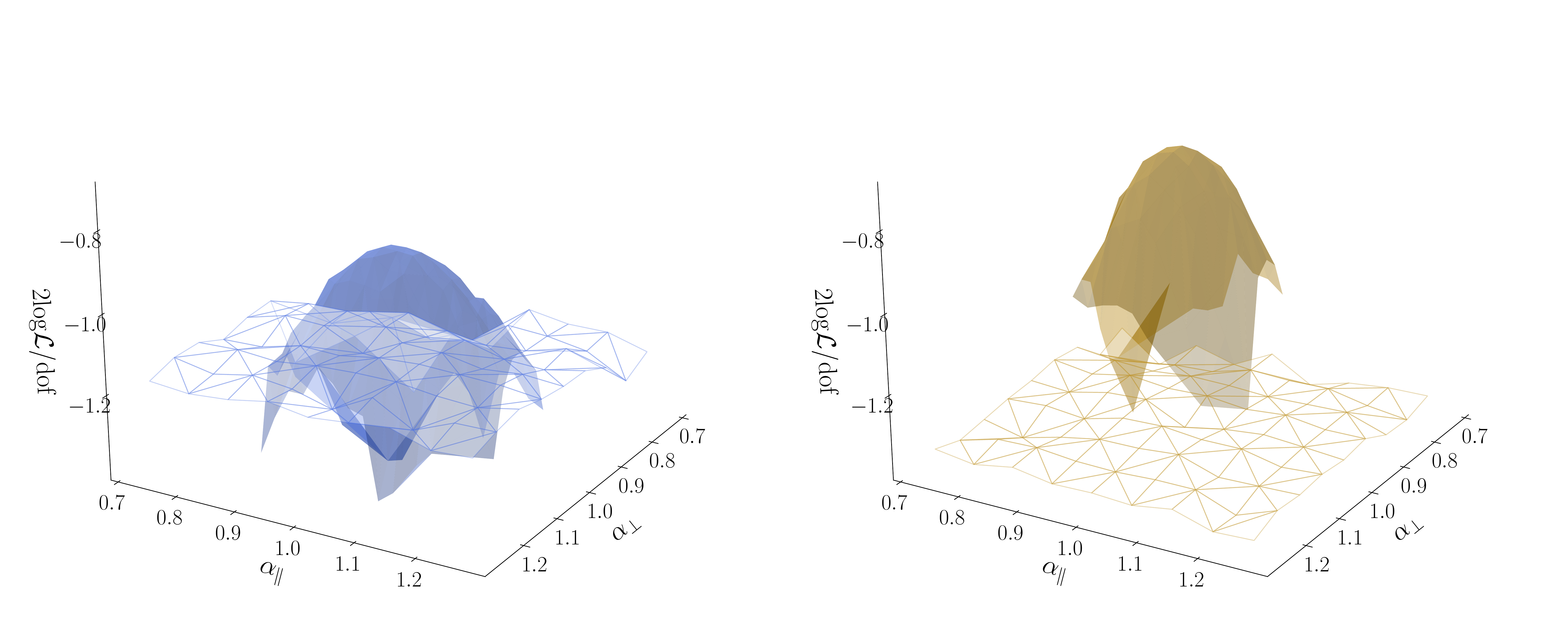}
\caption{Two-dimensional marginalized maximum likelihood surfaces of $\alpha_\perp$, $\alpha_\parallel$, before (left) and after (right) eFAM reconstruction. Solid surfaces describe the best-fit model including the acoustic feature, while transparent surfaces depict the best-fit model without BAO \citep{EisensteinHu1998}. }\label{fig:Data_significance}
\end{figure*}

In this section, we apply eFAM to the SDSS DR12 galaxy sample and analyze the clustering properties before and after eFAM reconstruction. 

To perform the reconstruction, we set the fiducial cosmology to the one of the mocks. We weigh the mass of each galaxy by its linear bias, $b=b(z_\mathrm{gal})$, estimated interpolating the redshift-to-bias relation of SDSS-III galaxies at the redshift of the galaxy, $z_\mathrm{gal}$ \citep{Bz}.  

For the present study, we consider the Obs and RecL sample. Here, distances are assigned to redshift coordinates using the same (fiducial) cosmological of reconstruction. To avoid to be affected by spurious tidal fields induced by edge effects, we exclude from the analysis all galaxies lying in a buffer region within $d_\mathrm{Buffer}=125h^{-1}\mathrm{Mpc}$ from the sample's edges.

For the two samples, we evaluate the 2PCF of the galaxies using the statistical weights described in Section~\ref{subsec:data} and measure its its parallel and perpendicular clustering wedges. Finally, we compare $\xi_\perp$ and $\xi_\parallel$ with the models presented in Section~\ref{sec:ximodel}.

Similarly to Ross17, we focus the analysis on the estimate of $H(r_\mathrm{s}/r_\mathrm{s}^\mathrm{fid})$ and $D_\mathrm{A}(r_\mathrm{s}^\mathrm{fid}/r_\mathrm{s})$, largely determined from the modelling of the BAO peak. We follow the same procedure adopted in Section~\ref{subsec:AcoustiMocks}. We model the clustering wedges in the separation range $50<s<150$ and fix the values of $f$, $\Sigma_\mathrm{s}$, $\Sigma_\parallel$, and $\Sigma_\perp$ to those measured in the mocks (Table~\ref{tab:MDPatchy_alpha}). In particular, for the RecL sample, we set $f=\Sigma_\mathrm{s}=0$ after having verified that the effective removal of RSD in the reconstructed data.

In the likelihood analysis, we use the covariance matrix estimated from the mocks, as described in Section~\ref{subsec:cov}. We note here that, although eFAM reconstruction is rather insensitive to the background cosmology \citep{nusser1998omega}, a mismatch between the fiducial and correct values of the cosmological parameters may generate spurious peculiar velocities and residual anisotropies in the reconstructed 2PCF of the SDSS-DR12 galaxies.
We search for such anisotropies by comparing the two clustering wedges and proceed with the likelihood analysis only after having excluded their presence.

We illustrate the results of our analysis in Figure~\ref{fig:Data_wedges}. Top panel shows the SDSS-DR12 parallel and perpendicular wedge as measured in redshift-space before eFAM reconstruction. The bottom panel shows the same quantities measured after the reconstruction. Red-triangles and blue-dots show $\xi_\parallel$ and $\xi_\perp$, respectively, multiplied by the square of the spatial separation, $s^2$. The error-bars are the 1$\sigma$ uncertainties estimated from the mocks, which corresponds to the diagonal elements of the covariance matrix. The BAO peak is clearly seen in all measurements. To highlight its statistical significance, we over-plot the best fitting models (Equation~\ref{eq:xi_temp}) represented by continuous and dashed, solid curves. Dotted and dot-dashed curves show the best fit clustering wedge model with no BAO feature in it, obtained using the no-wiggle power spectrum $P_\mathrm{nw}$ of \cite{EisensteinHu1998}.

As expected, in redshift-space (top panel) $\xi_\parallel$ and $\xi_\perp$ are significantly different from each other because of RSD. 
In the RecL sample (bottom panel), the two wedges are in good agreement, confirming the efficiency of the eFAM at recovering the real-space correlation function and thus justifying our decision to set $f=\Sigma_\mathrm{s}=0$ in the model fitting.
From the plots, it is also evident that the eFAM reconstruction increases the amplitude and the sharpness of the acoustic peak. 

The results of the likelihood analysis corroborate and quantify this visual impression.
Figure \ref{fig:Data_significance} show the two-dimensional marginalised maximum likelihood of the dilation parameters $\alpha_\perp$, $\alpha_\parallel$ surfaces before (left panel) and after eFAM reconstruction (right panel). 
Solid surfaces show the likelihood of models that include the BAO peak while transparent surfaces depict the case in which no acoustic feature is included in the model.
The comparison between the amplitudes of the peaks in the solid and transparent likelihood surfaces is a proxy of the significance of the BAO peak detection. 

\begin{table}
	\centering
	\caption{Fit results from the SDSS-DR12 Combined Sample.}
	\label{tab:DataResults}
	\resizebox{1.\columnwidth}{!}{%
	\begin{tabular}{lcc}
		\hline
		sample &\begin{tabular}{@{}c@{}}$D_\mathrm{A}(0.38)(r^\mathrm{fid}_\mathrm{s}/r_\mathrm{s})$\\  $\mathrm{(Mpc)}$\end{tabular} & \begin{tabular}{@{}c@{}}$H(0.38)(r_\mathrm{s}/r^\mathrm{fid}_\mathrm{s})$\\ (km~s$^{-1}$~Mpc$^{-1}$)\end{tabular}\\
		\hline
		
		Obs &1129  $\pm$ 40 &82 $\pm$  4 \\
		
		
	    RecL & 1090 $\pm$ 29 &83 $\pm$ 3\\
	    \bottomrule
	 \end{tabular}
	 }
\end{table}

Before reconstruction, non-linear RSD undermine the agreement with the fiducial model yielding a modest maximum in the likelihood. Alongside, the limited prominence of the acoustic feature legitimates the model without acoustic oscillations. For eight free fitting parameters, the maximum $\Delta\chi^2$ between the two surfaces translates into a $2.1\sigma$ BAO detection. After reconstruction, the accurate modelling of both RSD and non-linear clustering favours the fiducial template against the no-wiggle model rising the significance of the BAO detection to $4.4\sigma$.

Let us now focus on the estimate of the two parameters $H$ and $D_\mathrm{A}$ derived from the dilation parameters. Equations~(\ref{eq:alpha_perp}, \ref{eq:alpha_par})
Figure~\ref{fig:Data_cont} shows the probability contours level of the bi-variate Gaussian that best fits the marginalized likelihood.
Contours and shaded areas represent the 1, 2 and 3$\sigma$ confidence levels of the Gaussian before (blue) and after (yellow) eFAM reconstructions. Best fitting values are shown with a blue circle and yellow star, respectively. In the likelihood analysis the two parameters 
$H(r_\mathrm{s}/r_\mathrm{s}^\mathrm{fid})$ and $D_\mathrm{A}(r_\mathrm{s}^\mathrm{fid}/r_\mathrm{s})$,
are measured at the redshift of the sample $z=0.38$. The corresponding fiducial values are
$H^\mathrm{fid}=83(\mathrm{km/s})^{-1}(\mathrm{Mpc})^{-1}$ and $D^\mathrm{fid}_\mathrm{A}=1098\mathrm{Mpc}$.

Table~\ref{tab:DataResults} quantifies these considerations. The best fit value of $D_\mathrm{A}$ is shifted by $\sim 3.5$ per cent, comparably to what seen in the analysis of the mock catalogues (Table~\ref{tab:alpha_alpha}. There, the effect of eFAM was to increase the magnitude of $\alpha_\parallel$, bringing it into an agreement with the expected value. We thus assume that, similarly, eFAM successfully removes systematic errors on $\alpha_\parallel$, and hence on $D_\mathrm{A}$, from the real dataset too. In support of this hypothesis, we note that Ross17 (see Table~3 in \cite{Alam:2016hwk} where $D_\mathrm{M}=D_\mathrm{A}(1-z)$) found $H(r_\mathrm{s}/r^\mathrm{fid}_\mathrm{s)}=(81\pm2)$ (km~s$^{-1}$~Mpc$^{-1}$) and $D_\mathrm{A}(r^\mathrm{s}_\mathrm{s}/r_\mathrm{s})=(1096\pm17)\,\mathrm{Mpc}$ in their post-reconstruction analysis, which are in best agreement with our post-reconstruction measurements. 
On top of this, eFAM also reduces random errors. Relative errors on $D_\mathrm{A}$ decrease from $\sigma_{D_\mathrm{A}}/D_\mathrm{A}\simeq 3.5$ per cent, to $\simeq 2.6$ per cent and statistical uncertainties on $H$ decrease from $\sigma_{H}/H \simeq 5$ per cent to $\simeq 3.6$ per cent.
We further discuss the amplitudes of the parameters uncertainties in Section~\ref{sec:Discussion}.

\vspace{-0.4cm}
\section{Discussion and Conclusions}\label{sec:Discussion}

In this study, we have applied the eFAM reconstruction algorithm to a subset of the SDSS-DR12 Combined Sample of galaxies aiming at assessing the performances of eFAM when applied to a real dataset characterized by specific selection effects and observational uncertainties.

This work is the follow-up of the study conducted in \cite{Sarpa:2018ucb}. There, we have appraised the performance of eFAM using a somewhat idealized, simulated datasets with $\mathcal{O}(10^6)$ objects, similar in size to that of current and future spectroscopic redshift surveys. The results of our past analysis have shown that the first implementation of eFAM was indeed able to successfully reconstruct the real-space positions and velocities of each object at any epochs, and specifically at both the observed redshift and at a back-in-time epoch in which density fluctuations were still evolving in the linear regime. Urged by the rising need for accurate reconstruction techniques able to enhance the signal-to-noise of the acoustic feature in the two-point correlation function, we have focused or study on the BAO scale. Despite being of high scientific interest, the application of eFAM to the mere study of the BAO scale is a bit of a limitation. Indeed, eFAM is a non-perturbative, non-linear technique designed to describe non-linear self-gravitating systems,  and thus it is able to extract information from scales much smaller than the BAO one.

In the current study, we have repeated the original analysis using a real dataset and a suite of realistic simulated catalogues. We chose the  SDSS-DR12 Combined Sample for three reasons. First of all, its size and characteristics are representative of the state-of-the-art surveys as well as next-generation, wide surveys such as DESI or Euclid. Second, reconstruction techniques based on the Zel'dovich approximation have been already applied to this dataset, mainly to extract scientific information from the position and the amplitude of the BAO peak in the two-point correlation function, providing us with a reference for our results. When performing the comparison, we must keep in mind that this constitutes a specific test, limited to quasi-linear scales, where eFAM is supposed to perform just as well as Zel'dovich-based reconstruction algorithms. Third, our choice was encouraged by the public availability of a sufficiently large number of realistic mock catalogues mimicking its clustering properties. A large number of mocks is crucial to both calibrate our reconstruction technique to the specific characteristics of the survey and to estimate errors and their covariance.

To perform this analysis, we modified the eFAM algorithm to be able to deal with all the specific properties of real datasets that are not necessarily present in simulated catalogues. 
The first one is the galaxy bias. The first implementation of eFAM implicitly assumes that all the mass in the system is associated with discrete, visible objects than maintain their identity when traced back in time. \cite{Nusser+1991} have proposed a method to modify this assumption to include the linear bias. Here, we adopted an alternative approach. We associated with each galaxy a statistical weigh accounting for both the clustered dark matter component located at the galaxy position and the smooth dark component filling the survey volume. The inclusion of the smooth dark matter field is crucial to match the average mass density within the survey with the cosmic mean and hence to avoid the observed system to be seen as a local under-density.
Although we did not perform specific tests to evaluate the goodness of this approach, we did not find any evidence of detectable systematic errors introduced by this biasing scheme in any of the validation tests we performed.

The second aspect we considered in the new implementation of eFAM is the gravitational influence of the unknown mass distribution lying outside the volume of the survey. The problem is general. The lack of clustering information outside the survey volume prevents us to fully reproduce the bulk motions within the survey, while the gravitational potential estimated from a finite and possibly anisoropic volume induces deceptive tidal fields in the proximity of the edges of the survey. In \citet{Sarpa:2018ucb}, we were able to minimize those effects by considering a spherical geometry and discarding from the post-reconstruction analysis all the objects lying in the proximity of the spherical surface. However, real surveys are often characterized by more complicated geometry, hence modelling the influence of external matter distribution for real datasets is a serious problem. We proposed here a solution consisting of a two-step procedure. First, we modelled the tidal field and assess their impact under the assumption of a homogeneous isotropic mass distribution outside the sample. Second, as in our original work, we defined a buffer region near the survey's edges. Objects in the buffer region are accounted for in the eFAM reconstruction but excluded from scientific analyses.
We run several dedicated tests to validate this procedure and chose the size and shape of the buffer region such as to optimize the compromise between a meaningful statistical sampling and systematic uncertainties induced by tidal effects.
We proceeded as follows. eFAM is supposed to displace objects from their observed redshift to their back-in-time real-space position. When it does it successfully, it removes all the anisotropies in the two-point correlation function due to RSD. In our tests, we found that an imperfect removal of tidal effects would manifest itself as a non-zero quadrupole moment on scales larger than the BAO peak. One can then enlarge the buffer region until this spurious signal is completely removed.
The availability of realistic mock catalogues is obviously of paramount importance to calibrate the procedure. 
It is worth stressing that tidal field will become an even more relevant issue with the advent of next-generation wide surveys. Indeed, to minimize evolution effects, typical sub-samples used for scientific analyses will have a depth along the radial direction much shorter than the their transverse size.
The impact of the tidal field will thus play an important role in optimizing the shape and the volume of the sample to be analyzed.

A distinctive trait of eFAM reconstruction is its ability to reconstruct the full orbit of the mass tracers, as opposed to solely provide their positions at some fixed early epoch.
We can thus use eFAM to recover the real-space positions of objects at the observed and high redshift, alike, effectively modelling both redshift distortions and non-linear evolution.
This is done by assuming a value for the linear growth rate $f$. The success of the reconstruction can be then assessed by searching for residual anisotropy in the real-space two-point correlation function of the galaxies.
Our tests on the mock reveal a remarkably accurate modelling of RSD. The values of $f$ and  $\Sigma_\mathrm{s}$ in the real-space 2PCF measured at the redshift of the sample are consistent with zero. Besides, the eFAM reconstruction reduces by 65 per cent the damping of the acoustic oscillations due to late-time non-linear clustering, $\Sigma_{\parallel,\perp}$.
Having successfully removed RSD, we can set the parameters $f$ and  $\Sigma_\mathrm{s}$ equal to zero when modelling the 2PCF of galaxies at their back-in-time reconstructed positions, effectively decreasing the number of free parameters in the fit.
Furthermore, we found no need to include the additional shape parameters $B_{0,2}$ used in Ross17 to fit the 2PCF.
As a result, the back-in-time eFAM reconstructed two-point correlation function can be well fit with an 11-parameter model over the range $[50,150]h^{-1}$Mpc, to be compared with the 15-parameter model used to fit the measured 2-point correlation function in redshift-space.

A final, positive characteristic of eFAM worth stressing is its non-linear nature which allows us to push the analyses down to scales that are not accessible to Zel'dovich-based reconstructions.
This fact is already evident from the wide separation range, $[50,150]h^{-1}$Mpc, used in our analysis. However, our validation tests show that one can safely push the analysis down to separations as small as $25h^{-1}$Mpc without introducing systematic errors. Strictly speaking, in this work we only checked this to be true for the estimate of the dilation parameters. However, the visual inspections of the clustering wedges reveal no obvious departures from model predictions on those scales. In fact, we found that RSD are successfully removed and models fit the reconstructed 2PCF up to $180h^{-1}$Mpc.

We conclude by summarizing the main results of our analysis:

\begin{itemize}
    \item The tests performed with 400 SDSS-DR12 mock catalogs show that eFAM successfully reconstruct the real-space position of galaxies at the mean redshift of the sample, $z=0.38$. As a result, redshift-space distortions are removed from the two-point correlation signal in the separation range $[50,180]h^{-1}$Mpc. On larger scales, we detect non-zero quadrupole moments that we attribute to the gravitational pull of the inhomogeneous mass distribution outside the survey volume. The impact of these external tidal field can be reduced by excluding objects near the edge of the survey. The size of this exclusion region should be set by compromising between number of objects and range of separation to be used in the clustering analysis. Ultimately it depends on the volume, shape, and redshift of the sample. The lower bound in separation, set to $50h^{-1}$Mpc, simply represents a conservative choice. In fact, we showed that it can be safely reduced to $25h^{-1}$Mpc without compromising our ability to estimate the dilation parameters, $\alpha_{\perp,\parallel}$, and, therefore, the values of $H$ and $D_\mathrm{A}$.
     
    \item eFAM reconstruction allows us to increase the precision of both $\alpha_{\perp,\parallel}$ by $\sim 3$ per cent. This is a significant reduction which, however, is smaller in magnitude than the one obtained by Ross17 using the Zel'dovich approximation to perform the back-in-time reconstruction.
    This is somewhat surprising since one would expect, based on the results of \cite{Sarpa:2018ucb}, which the quality of the eFAM reconstruction should be higher than that of Zel'dovich approximation. Part of this discrepancy can be explained by the very nature of the analysis. Here, we are focusing on the $\alpha_{\perp,\parallel}$ parameters that are mainly determined by the reconstruction quality of the BAO peak, and we do not eFAM to perform significantly better than Zel'dovich on these scales.
    However, we do not expect worse performances either, especially considering that the number of free parameters used in our model for the reconstructed 2PCF is smaller than the one employed in Ross17.
    We believe that the larger eFAM uncertainty, that are estimated from the mock catalogs, reflect the fewer catalogs (400) and reduced volume (north galactic cap) used to evaluate the covariance matrix with respect to that (1000, both north and south caps) used by Ross17. 
    To corroborate this hypothesis, we notice that \cite{kazin2013clustering} performed a similar analysis estimating the values of the dilation parameters and their uncertainties using 600 rather than 1000 mocks. The uncertainties presented in their work, $\sigma_H/H=5.8$ per cent and $\sigma_{D_\mathrm{A}}/D_\mathrm{A}=3.1$ per cent (see Table~3), are overestimated with respect Ross17 and comparable with ours.
     
    \item From the clustering wedges of the back-in-time reconstructed position of the SDSS-DR12 galaxies, we have estimated the Hubble parameter and the angular diameter distance at the effective redshift of the sample, normalized to their values in the fiducial cosmological model. We found $D_\mathrm{A}(r^\mathrm{fid}_\mathrm{s}/r_\mathrm{s})=1090 \pm 29 $ Mpc and $H(r_\mathrm{s}/r^\mathrm{fid}_\mathrm{s})= 83 \pm 3$(km~s$^{-1}$~Mpc$^{-1}$).
    The eFAM reconstruction significantly reduces the error in these estimates by 35 and 39 per cent, respectively, compared to the case in which the analysis is carried out in redshift-space, i.e. before reconstruction.
    These results are in good agreement with those of previous, similar analyses performed on the same datasets (Ross17,\cite{Vargas-Magana:2016imr}).

\end{itemize} 

The results presented in this work show that eFAM can be successfully applied to current and future spectroscopic datasets with 
$\mathcal{O}(10^6)$ objects. The enhancement of the clustering signal-to-noise at the BAO peak is comparable to that obtained by the popular reconstruction techniques that assume Zel'dovich approximation.
However, eFAM also succeeds in recovering the correlation signal well into non-linear regime, as demonstrated by its ability to recover the correct shape and amplitude of the galaxy-galaxy correlation function down to separation as small as $25 h^{-1}$Mpc.
We conclude that eFAM provides a back-in-time reconstruction method that, although not quite as fast as those based on Zel'dovich approximation, can be regarded as a complementary tool to extract scientific information from the numerous small-scale modes that cannot be accessed by standard reconstruction techniques.

\vspace{-0.6cm}
\section*{Acknowledgements}
ES thanks A. Longobardi for valuable discussions and precious scientific advice.

CS and ES acknowledge financial support by Programme National Cosmology et Galaxies (PNCG) of CNRS/INSU with INP and IN2P3, co-funded by CEA and CNES. 
EB is supported by MUIR/PRIN 2017 “From Darklight to Dark Matter: understanding the galaxy-matter connection to measure the Universe”,  ASI/INAF agreement n. 2018-23-HH.0 “Scientific activity for Euclid mission, Phase D”, ASI/INAF agreement  n. 2017-14-H.O “Unveiling Dark Matter and Missing Baryons in the high-energy sky”  and INFN project “INDARK”.
This research was supported by the Munich Institute for Astro- and Particle Physics (MIAPP) which is funded by the Deutsche Forschungsgemeinschaft (DFG, German Research Foundation) under Germany´s Excellence Strategy – EXC-2094 – 390783311 and was also partially supported by the Programme National Cosmology et Galaxies (PNCG) of CNRS/INSU with INP and IN2P3, co-funded by CEA and CNES

The massive production of all MultiDark-Patchy mocks for the BOSS Final Data Release has been performed at the BSC Marenostrum supercomputer, the Hydra cluster at the Instituto de Fısica Teorica UAM/CSIC, and NERSC at the Lawrence Berkeley National Laboratory. We acknowledge support from the Spanish MICINNs Consolider-Ingenio 2010 Programme under grant MultiDark CSD2009-00064, MINECO Centro de Excelencia Severo Ochoa Programme under grant SEV- 2012-0249, and grant AYA2014-60641-C2-1-P. The MultiDark-Patchy mocks was an effort led from the IFT UAM-CSIC by F. Prada's group (C.-H. Chuang, S. Rodriguez-Torres and C. Scoccola) in collaboration with C. Zhao (Tsinghua U.), F.-S. Kitaura (AIP), A. Klypin (NMSU), G. Yepes (UAM), and the BOSS galaxy clustering working group.




\bibliographystyle{mnras}
\bibliography{biblio} 



\appendix

\vspace{-0.5cm}
\section{Correction for survey geometry: accuracy test}\label{appendix:Survey_geom}
\begin{figure}
\centering
\includegraphics[width=1\columnwidth]{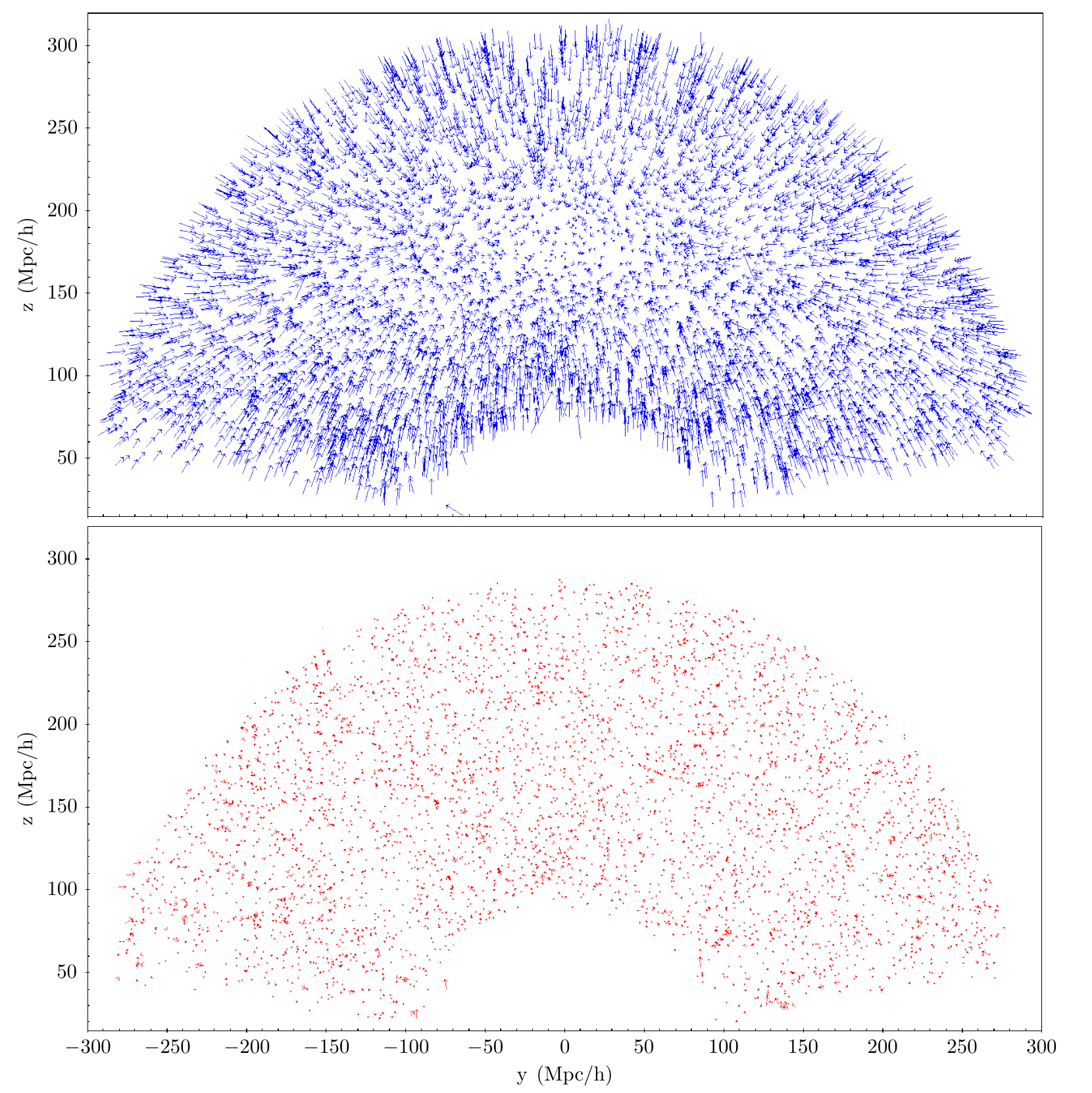} 
\caption{2D velocity map for the BOSS-like random distribution.
\emph{Top panel} velocity field reconstructed without correcting the external tidal field; the bulk-flow induced by the shape of the survey is dominant. \emph{Bottompanel:} velocity field reconstructed including the external tidal field; the geometrical bulk flow is successfully removed.}
\label{fig:Correction_random}
\end{figure}

To asses the validity of the procedure described above, we applied eFAM on a uniform random catalogue with same sky-coverage and redshift proportions of the SDSS-DR12 survey. Figure~\ref{fig:Correction_random} depicts the reconstructed velocity field obtained neglecting the external contribution (top panel) and including the correction for the survey geometry (bottom panel). As expected, in the first case the velocity field is dominated by a spurious coherent bulk-flow generated by the non-spherically symmetry of the survey. Adopting the correct estimation for $\phi_\mathrm{ex}$, the reconstructed velocity field is instead characterized by almost vanishing and randomly oriented peculiar velocities. This illustrates the efficiency of the correction in erasing the spurious motions and, more importantly, reassuring from the presence of additional systematics. 

\begin{figure}
\centering
\includegraphics[width=1\columnwidth]{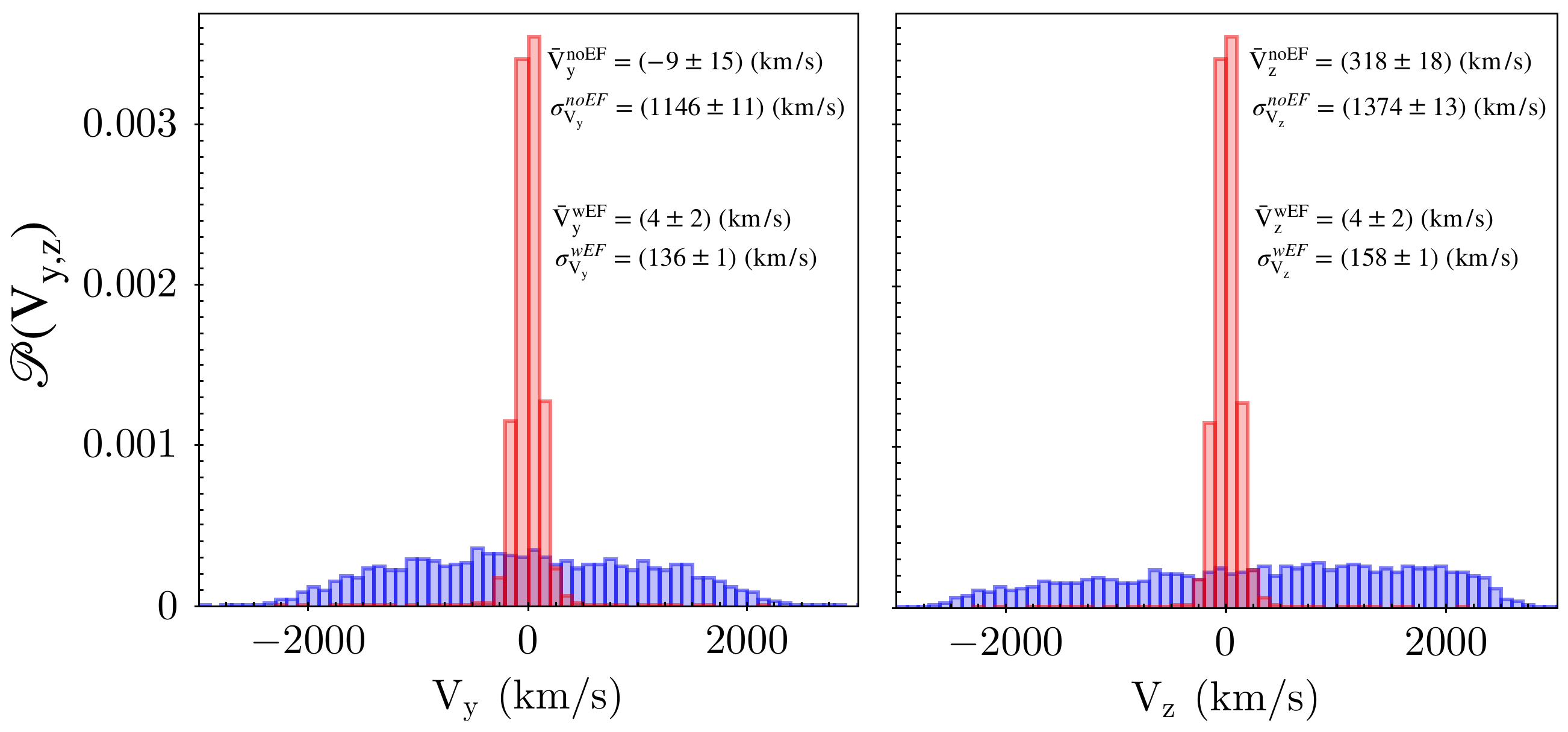} 
\caption[Probability distribution function of two components of the reconstructed peculiar velocities for the BOSS-like random distribution.]{Probability distribution function of two components of the reconstructed peculiar velocities for the BOSS-like random distribution. Before accounting for the external tidal field (top-right) the velocity distribution is characterized by a strong velocity dispersion of about $\sim 1400$ km/s, describing the spurious in-fall of particles induced by the survey geometry. With the inclusion of the external field (bottom-left) the rms of the velocities is of about $\sim 140$ km/s, proving the efficiency of the method in removing the bulk-flow.}
\label{fig:PDF_Correction_random}
\end{figure}

A more quantitative description of the efficiency of the correction is provided by Figure~\ref{fig:PDF_Correction_random}, which illustrates the PDF of two components of the peculiar velocities. Before applying the correction (blue), the reconstructed PDF is characterized by a wide dispersion accounting for the artificial in-fall of particles and, for the $\mathrm{V}_z$ component, by a spurious bulk flow component. With the new method, the rms of the velocities is reduced by 90 per cent and the bulk flow removed, proving the efficacy of the method.

\begin{figure}
\centering
\includegraphics[width=1\columnwidth]{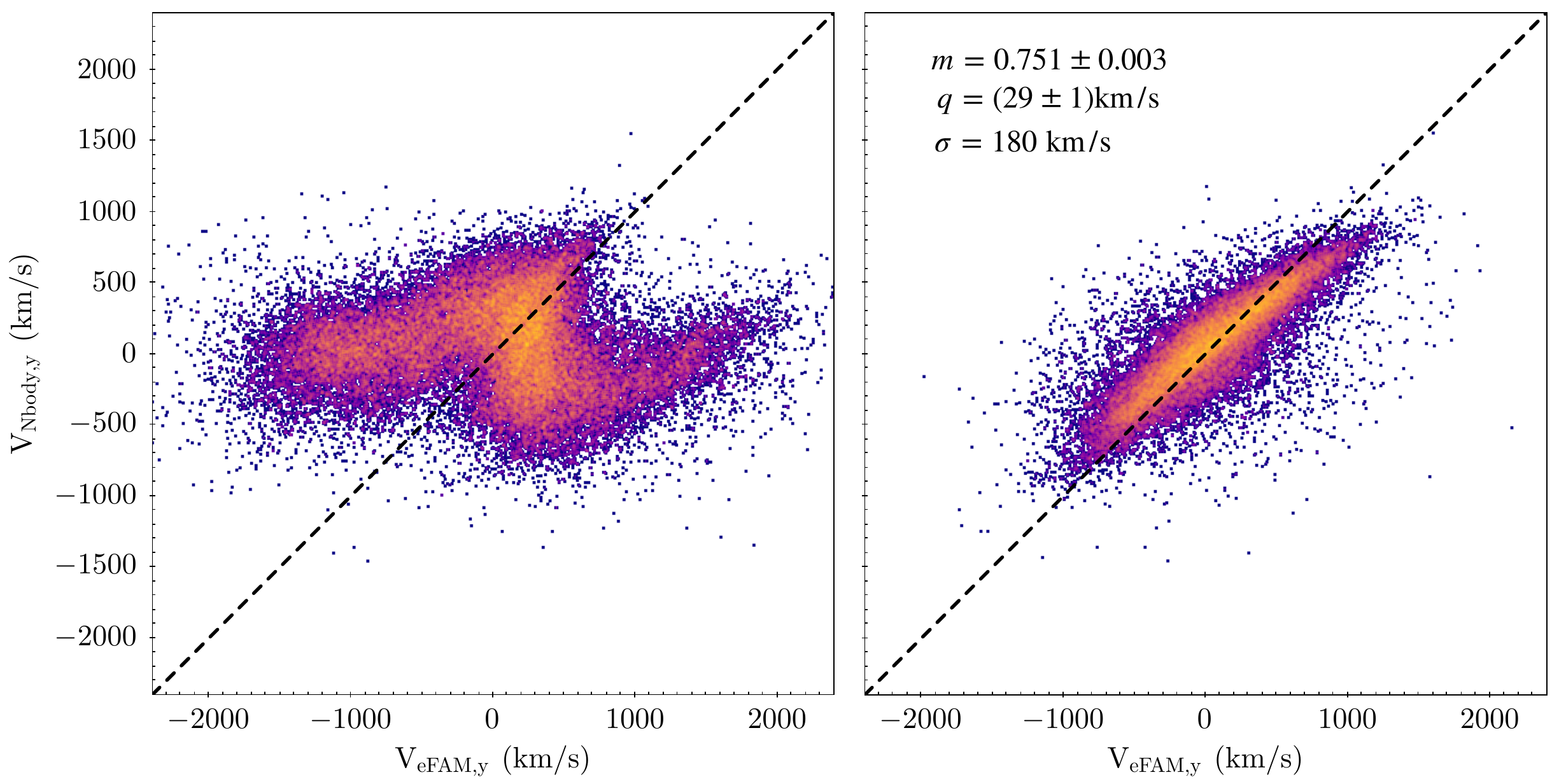}

\caption[Accuracy tests for eFAM velocity reconstruction applied on a non-spherically symmetric survey.]{Accuracy tests for eFAM velocity reconstruction applied on a non-spherically symmetric survey. ``True'' vs. reconstructed peculiar velocities for one Cartesian component before (left) and after (right) including the external tidal field computed from a random distribution (results are similar for other components). Along every Cartesian direction, a perfect reconstruction would give a linear regression $V_\mathrm{Nbody} = mV_\mathrm{eFAM} + q$ with slope $m = 1$ , no residual bulk velocity ($q = 0$), and no scatter (solid line). The reconstructed peculiar velocities before including the tidal field are not correlated with the true ones, having a significantly larger velocity dispersion with respect to the N-body velocities. After including the external tidal field the reconstructed peculiar velocities are instead well-correlated with the true ones.}\label{fig:tidal_scatter}
\end{figure}

The improvement in the velocity reconstruction due to the external field correction can be more quantitatively assessed through the velocity-velocity comparison illustrated in Figure~\ref{fig:tidal_scatter}. Before applying the geometrical correction (left panel), the reconstructed velocities are completely uncorrelated with the ``true'' ones, showing high-velocity tails which don't find correspondence in the $N$-body field. The correlation is efficiently restored after including the external field (right panel), yielding a precise reconstruction.

\bsp	
\label{lastpage}

\end{document}